\RequirePackage{booktabs}
\let\OriginalToprule\toprule
\let\OriginalMidrule\midrule
\let\OriginalBottomrule\bottomrule
\documentclass{interact}
\let\toprule\OriginalToprule
\let\midrule\OriginalMidrule
\let\bottomrule\OriginalBottomrule

\usepackage[T1]{fontenc}
\usepackage[utf8]{inputenc}
\DeclareUnicodeCharacter{1EC5}{\~{\^e}}
\usepackage{lmodern}
\usepackage{microtype}
\usepackage{setspace}
\usepackage{amsmath,amssymb,mathtools}
\usepackage{booktabs}
\usepackage{graphicx}
\usepackage{enumitem}
\usepackage{array,longtable,multirow}
\usepackage{siunitx}
\usepackage{pdflscape}
\usepackage{float}
\usepackage{placeins}
\usepackage{algorithm}
\usepackage{algpseudocode}
\usepackage{tikz}
\usetikzlibrary{arrows.meta,positioning,shapes.geometric}
\usepackage{xcolor}
\usepackage{natbib}
\bibpunct[, ]{(}{)}{;}{a}{}{,}
\renewcommand\bibfont{\fontsize{10}{12}\selectfont}
\usepackage[hidelinks,hypertexnames=false]{hyperref}
\usepackage{orcidlink}
\AtBeginDocument{%
  \setlength{\heavyrulewidth}{.08em}%
  \setlength{\lightrulewidth}{.05em}%
  \setlength{\cmidrulewidth}{.03em}%
  \setlength{\belowrulesep}{.65ex}%
  \setlength{\belowbottomsep}{0pt}%
  \setlength{\aboverulesep}{.4ex}%
  \setlength{\abovetopsep}{0pt}%
  \setlength{\cmidrulesep}{\doublerulesep}%
  \setlength{\cmidrulekern}{.5em}%
  \setlength{\defaultaddspace}{.5em}%
}
\makeatletter
\long\def\LT@makecaption#1#2#3{%
  \LT@mcol\LT@cols c{\hbox to\z@{\hss\parbox[t]\LTcapwidth{%
    \reset@font\fontsize{8}{10}\selectfont
    \sbox\@tempboxa{#1{{\bfseries\tablename\nobreakspace\thetable.}\hskip7pt}#3}%
    \ifdim\wd\@tempboxa>\hsize
      #1{{\bfseries\tablename\nobreakspace\thetable.}\hskip7pt}#3%
    \else
      \hbox to\hsize{\hfil\box\@tempboxa\hfil}%
    \fi
    \endgraf\vskip\baselineskip}%
  \hss}}}
\makeatother

\newcommand{\PSCALB}{\textsc{PSCALB3PM}}
\newcommand{\SALB}{\textsc{SALB3PM}}

\newcommand{\MPSC}{\ensuremath{M^{PSC}}}
\newcommand{\MSLS}{\ensuremath{\mathrm{MS}\!\times\!\mathrm{LS}^{PSC}}}

\newcommand{\tabmissing}{\multicolumn{1}{c}{--}}
\newcommand{\tabna}{\multicolumn{1}{c}{n/a}}

\begin{document}

\title{SAT Solving for Power Peak Minimization in the Parallel --Serial--with--Crossover Assembly Line Balancing Problem}
\author{
\name{Phong Nguyen Chi\orcidlink{0009-0007-0364-2219}\textsuperscript{1} and Khanh To Van\orcidlink{0009-0008-1907-7848}\textsuperscript{1}*\thanks{Corresponding author. E-mails: \texttt{khanhtv@vnu.edu.vn}}}
\affil{\textsuperscript{1}Faculty of Information Technology,\\VNU University of Engineering and Technology, Hanoi, Vietnam}
}
\history{Preprint}
\maketitle
\enlargethispage{11pt}

\begin{abstract}
The Parallel--Serial-with-Crossover Assembly Line Balancing Problem with Power Peak Minimization (PSCALB3PM) was introduced by Delorme and Gianessi (2022), who proposed an integer linear programming (ILP) model, and further developed in 2024 with a multi-start local-search matheuristic. This paper presents an exact approach based on Boolean satisfiability (SAT) that combines a compact encoding with incremental SAT solving. Specifically, it develops a compact encoding of task-to-workstation precedence constraints and an incremental SAT-based optimization procedure for power peak minimization. We evaluate the proposed method on 375 benchmark instances and compare it with both reference methods. SAT finds feasible solutions for all 375 instances and certifies optimality for 159, compared with 346 feasible solutions and 126 optimality certificates obtained by the ILP method. Relative to the reference matheuristic, SAT achieves lower power peaks on 130 instances, whereas the matheuristic obtains lower values on only four. These findings establish SAT as a competitive exact method for PSCALB3PM, achieving higher feasible-solution coverage and more optimality proofs than the reference ILP model, while generally outperforming the existing matheuristic in terms of solution quality.
\end{abstract}

\begin{keywords}
Reconfigurable manufacturing systems; Assembly-line balancing; Power peak minimization; Exact optimization; SAT solving.
\end{keywords}

\section{Introduction}

Assembly-line balancing (ALB) assigns production tasks to ordered workstations while satisfying technological precedence and cycle-time constraints. Beyond conventional objectives such as minimizing the number of workstations or cycle time, modern manufacturing systems increasingly consider equipment configuration, resource allocation, and energy-related objectives \citep{SchollBecker2006,BattaiaDolgui2013,BoysenEtAl2022}. In particular, minimizing peak power demand is distinct from minimizing total energy consumption: when task durations and power requirements are fixed, changing task assignments and starting times can reduce simultaneous power demand without necessarily reducing the energy consumed during production \citep{GahmEtAl2016,PetersenEtAl2025}. This distinction motivates joint task assignment and scheduling approaches that explicitly account for the temporal overlap of production activities.

Reconfigurable manufacturing systems provide flexibility through changes in system configuration and resource allocation \citep{KorenEtAl1999,KorenEtAl2017}. Among these systems, parallel--serial assembly lines with crossover allow alternative task flows and resource configurations, creating additional opportunities for balancing production and reducing peak power demand \citep{FreiheitEtAl2004}. \citet{DelormeGianessi2022} introduced the Parallel--Serial-with-Crossover Assembly Line Balancing Problem with Power Peak Minimization (PSCALB3PM) and proposed a time-indexed integer linear programming (ILP) model. In their extended study, \citet{DelormeGianessi2024} presented this formulation as \MPSC{} and developed a multi-start local-search matheuristic, \MSLS{}. PSCALB3PM jointly determines task assignments, task starting times, and the number of parallel resources activated at each workstation, subject to a prescribed cycle time, workstation-level resource limits, and a global resource budget. These decisions are tightly coupled: resource allocation determines the available time horizon at each workstation, while power demand must account for activities that occur at time points separated by integer multiples of the cycle time but correspond to the same phase of successive production cycles. Consequently, an effective formulation must integrate assignment, scheduling, resource feasibility, and cyclic power aggregation.

Boolean satisfiability (SAT) solving offers an alternative framework for exact combinatorial optimization, including scheduling and power-peak minimization. Previous SAT, maximum satisfiability (MaxSAT), and incremental SAT solving approaches have addressed the Simple Assembly Line Balancing Problem with Power Peak Minimization (SALB3PM)
\citep{PyEtAl2024,ZhengEtAl2024,ZhengEtAl2026,kieu2026compact}. The MaxSAT approach extends the encoding proposed by \cite{PyEtAl2024}, which employs Boolean decision variables for workstation assignments and task starting times. In contrast, \cite{kieu2026compact} develop a compact representation based on cumulative reachability variables and incremental SAT solving for optimizing peak-power bounds. However, extending these approaches to PSCALB3PM requires efficient SAT propagation that captures the allocation of multiple parallel resources at each workstation.
In particular, the encoding must link task assignments and starting times to resource-dependent workstation time horizons, enforce both workstation-order and within-workstation precedence constraints, and aggregate power demand across cycle-congruent time points. These requirements motivate a resource-aware SAT formulation tailored to the structure of PSCALB3PM.

This paper extends the compact SAT encoding proposed by \cite{kieu2026compact}, which demonstrated its effectiveness for SALB3PM, to PSCALB3PM and develops an exact SAT-based approach that combines this extended encoding with incremental SAT solving. The compact encoding uses cumulative reachability variables to represent assignment and starting-time decisions and to express precedence constraints without explicitly comparing every pair of workstation or time-index values. It is extended with resource-activation constraints, a global resource budget, and phase-based power constraints that account for cyclic production. For optimization, an initial feasible schedule provides an upper bound on the peak power, while successive tighter bounds are tested using the same SAT solver. This incremental strategy retains the feasibility formula and learned clauses across successive optimization steps, avoiding the need to rebuild and solve an independent SAT instance for each candidate bound.

The contributions of this paper are twofold.
First, we develop an exact, resource-aware SAT formulation for PSCALB3PM and establish its correspondence with the discrete-time scheduling model. The formulation integrates compact assignment and timing representations with resource-dependent time horizons and cyclic power aggregation, together with incremental SAT solving for minimizing the peak-power objective.
Second, we conduct a controlled computational evaluation on the 375 benchmark instances introduced by \cite{DelormeGianessi2024}. The evaluation compares the proposed SAT encoding with the SAT encoding of \cite{PyEtAl2024}, using the same incremental SAT solving design, as well as with the reference ILP model \MPSC{} and matheuristic \MSLS{} presented in \cite{DelormeGianessi2024}. This design distinguishes the effects of the Boolean representation from those of the optimization procedure and enables a consistent assessment of feasibility, solution quality (optimized power peak values), and optimality certification. SAT achieves broader feasible-solution coverage, proves more instances optimal, and delivers better power-peak solutions than the reference methods.

The remainder of the paper is organized as follows.
Section~\ref{sec:literature} reviews related work. Section~\ref{sec:problem} presents the problem definition and the reference ILP model. Section~\ref{sec:sat} describes the SAT formulations and incremental SAT solving approach for PSCALB3PM. Sections~\ref{sec:experiments} and~\ref{sec:results} present the computational protocol and experimental results, respectively. Section~\ref{sec:conclusion} concludes the paper. 

\section{Related work}\label{sec:literature}

Assembly-line balancing problems differ in their task assumptions, workstation capabilities, layout, and objectives. The surveys by \cite{SchollBecker2006}, \cite{BattaiaDolgui2013}, and \cite{BoysenEtAl2022} provide complementary classifications of these problem settings and their solution methods. Parallel-line balancing is itself a broad class: \citet{AguilarEtAl2020} classify settings in which several assembly lines are balanced together. Such settings should be distinguished from a parallel--serial line with multiple resources at individual workstations.

Within reconfigurable manufacturing, capacity and functionality are design decisions rather than fixed characteristics \citep{KorenEtAl1999}. \citet{KorenEtAl2017} examine design for scalability, while \citet{FreiheitEtAl2004} analyze productivity in paced parallel--serial configurations with and without crossover. At the balancing level, \citet{LahrichiEtAl2021} minimize the number of machines under a prescribed cycle time, considering parallel machines within workstations, sequence-dependent setups, and task-compatibility restrictions. Their split-based hybrid method solves the balancing subproblem exactly for a fixed overall task sequence. \citet{CerqueusDelorme2023} assess scalability at the design stage of reconfigurable manufacturing systems using a hypervolume indicator based on the takt times and resource requirements of reachable configurations. They use complete enumeration to examine how task allocation and permitted reconfiguration changes affect scalability. These studies demonstrate the importance of line architecture and resource allocation in production planning. However, they do not address the cyclic peak-power minimization objective considered in \PSCALB{}.

Energy-aware production planning encompasses total energy consumption, electricity cost, carbon emissions, and peak power demand \citep{GahmEtAl2016}. \citet{PetersenEtAl2025} review energy-related objectives and methods in assembly-line balancing. For robotic lines, \citet{MukundNilakantanEtAl2015} develop time-oriented and energy-oriented models that prioritize cycle time and total energy consumption in opposite orders. \citet{LiEtAl2016}, \citet{SunWangPeng2020}, and \citet{SoysalKurtIsleyen2022} minimize cycle time and total energy consumption jointly. For mixed-model parallel robotic lines, \citet{SoysalKurtEtAl2025} minimize joint cycle time and average energy consumption across production cycles. \citet{KubilayEtAl2026} formulate cycle-time-oriented and energy-consumption-oriented variants using constraint programming. \citet{CaoEtAl2026} jointly minimize cycle time and energy consumption using mixed-integer programming (MIP) with task-dependent robot operating modes. In contrast, peak-power minimization seeks to reduce the maximum aggregate instantaneous power demand by coordinating the timing and overlap of activities, even when task durations and power requirements are fixed. This distinction motivates scheduling formulations that explicitly model simultaneous power demand.

Research specifically addressing peak-power minimization in assembly lines has progressed from serial lines to more complex reconfigurable architectures. For serial assembly-line balancing, \cite{GianessiEtAl2019} introduce the Simple Assembly Line Balancing Problem with Power Peak Minimization (SALB3PM) and formulate an ILP model that jointly determines task assignments and starting times. \cite{LamyEtAl2020} investigate an earliest-start variant using multi-start evolutionary local search, while \cite{DelormeEtAl2023} incorporate an ILP decoder into local search to optimally split a fixed task sequence among workstations. For reconfigurable parallel--serial lines, \cite{DelormeGianessi2022} introduce PSCALB3PM and propose a time-indexed ILP model. Their extended study \citep{DelormeGianessi2024} retains this formulation, denoted \MPSC{}, and develops a multi-start local-search matheuristic, \MSLS{}, together with a broader computational evaluation.

PSCALB3PM extends peak-power line balancing by jointly considering task assignments, task starting times, and the allocation of parallel resources at each workstation. The decisions must satisfy technological precedence constraints, a prescribed cycle time, workstation-level resource limits, and a global resource budget. Moreover, because the production line operates cyclically, activities occurring at time points separated by integer multiples of the cycle time contribute to the same phase of successive production cycles and must be accounted for in the peak-power calculation. The resulting problem couples resource allocation, scheduling feasibility, and cyclic power aggregation. The reference ILP model provides a mathematical-programming approach for optimization and optimality certification, whereas the matheuristic combines multi-start local search over precedence-feasible workstation assignments with a decoder that evaluates schedules and additional resource requirements \cite{DelormeGianessi2024}. These methods provide the principal reference approaches for evaluating exact methods for PSCALB3PM.

SAT supports combinatorial optimization through decision queries and incremental solving, with its effectiveness depending on the encoding design and optimization strategy.
Recent studies have applied SAT and MaxSAT directly to peak-power minimization in serial assembly-line balancing. \cite{PyEtAl2024} propose a SAT encoding based on Boolean decision variables for workstation assignments and task starting times, together with activity variables and clauses enforcing uniqueness and power-related conflicts. Their optimization procedure excludes incumbent activity combinations responsible for a non-improving peak. \cite{ZhengEtAl2024} formulate the same problem using MaxSAT. Extending this line of research, \cite{kieu2026compact} develop a compact encoding based on cumulative reachability variables and investigate several optimization strategies, including incremental optimization using ordered peak-bound indicators. \cite{ZhengEtAl2026} further investigate exact SAT and MaxSAT formulations on a larger set of serial assembly-line balancing instances.

Although these studies establish the potential of SAT and MaxSAT for peak-power line balancing, the reviewed formulations focus primarily on serial lines and do not directly capture the coupled resource-allocation and cyclic scheduling requirements of PSCALB3PM. Extending them to this problem requires an encoding that links task assignments and starting times to resource-dependent workstation time horizons, enforces workstation and task precedence constraints, respects a global resource budget, and aggregates power demand across cycle-congruent time points. This paper addresses these requirements by extending the compact encoding of \cite{kieu2026compact} to PSCALB3PM and combining it with incremental SAT-based peak-power optimization. The SAT representation of \cite{PyEtAl2024} is also evaluated using the same incremental optimization procedure, allowing the effect of the Boolean representation to be assessed separately from that of the optimization strategy. Both SAT approaches are compared with the reference ILP model and matheuristic of \cite{DelormeGianessi2024} on the established benchmark instances.

Table~\ref{tab:related-literature} summarizes the studies most relevant to this work, highlighting their problem settings, features, solution methods, and distinguishing objective.

\begingroup
\enlargethispage{\baselineskip}
\setlength{\intextsep}{4pt}
\begin{table}[H]
\centering
\caption{Related studies.}\label{tab:related-literature}
\begingroup
\itshape 
\small
\setlength{\tabcolsep}{4pt}
\renewcommand{\arraystretch}{1.00}
\newlength{\relatedworkwidth}
\setlength{\relatedworkwidth}{\dimexpr\linewidth-6\tabcolsep\relax}
\begin{tabular}{@{}>{\raggedright\arraybackslash}p{.22\relatedworkwidth}>{\raggedright\arraybackslash}p{.25\relatedworkwidth}>{\raggedright\arraybackslash}p{.29\relatedworkwidth}>{\raggedright\arraybackslash}p{.24\relatedworkwidth}@{}}
\toprule
\normalfont Study & \normalfont Setting, Features & \normalfont Method & \normalfont Objective\\
\midrule
\multicolumn{4}{@{}l}{\textit{Resource allocation and energy-aware line balancing}}\\*\midrule \itshape 
\citet{LahrichiEtAl2021} & Parallel machines & Hybrid metaheuristic & Number of machines\\
\addlinespace[1pt]
\citet{CerqueusDelorme2023} & Parallel--serial lines with crossover & Complete enumeration & Scalability\\
\addlinespace[1pt]
\citet{MukundNilakantanEtAl2015} & Robotic assembly lines & Particle swarm & Cycle time; total energy consumption\\
\addlinespace[1pt]
\citet{LiEtAl2016} & Two-sided robotic lines& MIP, simulated annealing & Cycle time; total energy consumption\\
\addlinespace[1pt]
\citet{SoysalKurtEtAl2025} & Mixed-model parallel lines & Genetic algorithm & Joint cycle time; average energy consumption\\
\addlinespace[1pt]
\citet{KubilayEtAl2026} & Parallel robotic lines & Constraint programming & Cycle time; energy consumption\\
\addlinespace[1pt]
\citet{CaoEtAl2026} & Robotic lines & MIP, genetic algorithm & Cycle time; energy consumption\\
\midrule
\multicolumn{4}{@{}l}{\textit{Exact and metaheuristic approaches for power peak minimization in ALB}}\\*\midrule
\citet{GianessiEtAl2019} & Serial lines & ILP & Peak power \\
\addlinespace[1pt]
\citet{LamyEtAl2020} & Serial lines & Evolutionary local search & Peak power\\
\addlinespace[1pt]
\citet{DelormeEtAl2023} & Serial lines& ILP decoder, local search & Peak power \\
\addlinespace[1pt]
\citet{DelormeGianessi2022} & Parallel--serial lines with crossover & ILP & Peak power \\
\addlinespace[1pt]
\citet{DelormeGianessi2024} & Parallel--serial lines with crossover & ILP (\MPSC{}), multi-start local search (\MSLS{}) & Peak power\\
\midrule
\multicolumn{4}{@{}l}{\textit{SAT solving optimizations for power peak minimization in ALB}}\\*\midrule
\citet{PyEtAl2024} & Serial lines& SAT & Peak power \\
\addlinespace[1pt]
\citet{ZhengEtAl2024} & Serial lines & MaxSAT & Peak power \\
\addlinespace[1pt]
\citet{ZhengEtAl2026} & Serial lines & SAT; MaxSAT & Peak power \\
\addlinespace[1pt]
\citet{kieu2026compact} & Serial lines & Incremental SAT & Peak power\\
\addlinespace[1pt]
\midrule
\normalfont \textbf{This work} & Parallel--serial lines with crossover & Incremental SAT & Peak power\\
\bottomrule
\end{tabular}
\endgroup
\end{table}
\endgroup

\section{Problem definition and ILP model}\label{sec:problem}

\subsection{Problem setting and objective}

Let $O=\{0,\ldots,n-1\}$ be the set of production tasks. Each task $j\in O$ has an integer processing time $t_j>0$ and an integer power requirement $w_j>0$. Technological precedence relations form a directed acyclic graph $G=(O,E)$, where $(i,j)\in E$, equivalently $i\prec j$, means that task $i$ must precede task $j$. The line contains at most $m$ ordered workstations ($m\ge2$), indexed by $M=\{0,\ldots,m-1\}$. Each workstation can use at most $r_{\max}$ identical parallel resources, while the total number of resources is limited by a global resource budget $R_{\max}$. The prescribed cycle time is $c$. Following the reference benchmark, this study considers $r_{\max}\in\{1,2,3\}$.

A solution jointly determines the workstation assignment and starting time $s_j$ of each task, together with the number $r_k$ of resources at each workstation $k$. These decisions must respect workstation order along each precedence arc and ensure that tasks assigned to the same workstation satisfy temporal precedence and do not overlap in its representative virtual schedule. Following \citet{DelormeGianessi2024}, the $r_k$ parallel resources repeat this schedule with period $r_kc$, shifted from one another by $c$, to process successive products. Consequently, the time horizon of workstation $k$ is $r_kc$, while the production cycle remains $c$. The augmented timespan used for time-indexed variables is
\begin{equation}
H=r_{\max}c.
\tag{P-1}\label{eq:problem-horizon}
\end{equation}
We use the time-slot set $\mathcal{T}=\{0,\ldots,H-1\}$, the physical phase set $\mathcal{C}=\{0,\ldots,c-1\}$, and the candidate starting-time set
\begin{equation}
\mathcal{T}^{j}=\{0,\ldots,H-t_j\}.
\tag{P-2}\label{eq:problem-start-domain}
\end{equation}
For a workstation with $r$ resources, define $\mathcal{T}_{r}^{j}=\{0,\ldots,rc-t_j\}$ when $rc\ge t_j$, and $\mathcal{T}_{r}^{j}=\varnothing$ otherwise, including $\mathcal{T}_{0}^{j}=\varnothing$.

Because production is cyclic, activities at virtual time points separated by integer multiples of $c$ contribute to the same physical phase. Thus, the total power demand at phase $\tau$ aggregates activities at $\tau,c+\tau,\ldots,(r_{\max}-1)c+\tau$:
\begin{equation}
P_{\tau}=\sum_{j\in O}w_j\sum_{q=0}^{r_{\max}-1}
\mathbf{1}\{s_j\le qc+\tau<s_j+t_j\},
\qquad \tau\in\mathcal C.
\tag{P-3}\label{eq:problem-phase-power}
\end{equation}
The objective is to minimize the maximum power demand over one production cycle:
\begin{equation}
\min W_M,\qquad W_M=\max_{\tau\in\mathcal C}P_{\tau}.
\tag{P-4}\label{eq:problem-objective}
\end{equation}

\begin{figure}[!htb]
\centering
\includegraphics[width=.98\linewidth]{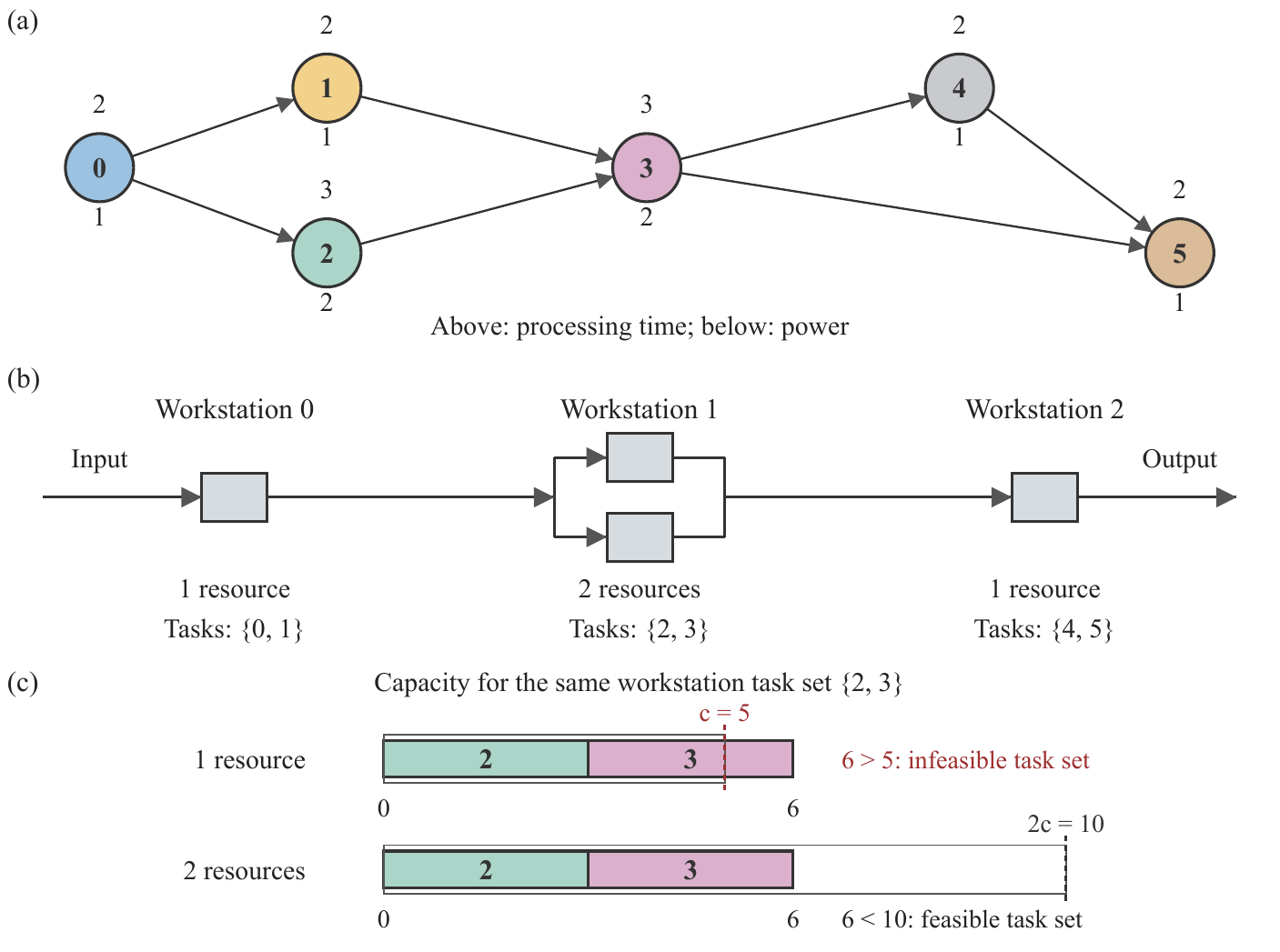}
\caption{Six-task example: (a) precedence and task data; (b) resource configuration with crossover; (c) virtual scheduling capacity with one and two resources.}
\label{fig:line-overview}
\end{figure}

\begin{figure}[H]
\centering
\includegraphics[width=.98\linewidth]{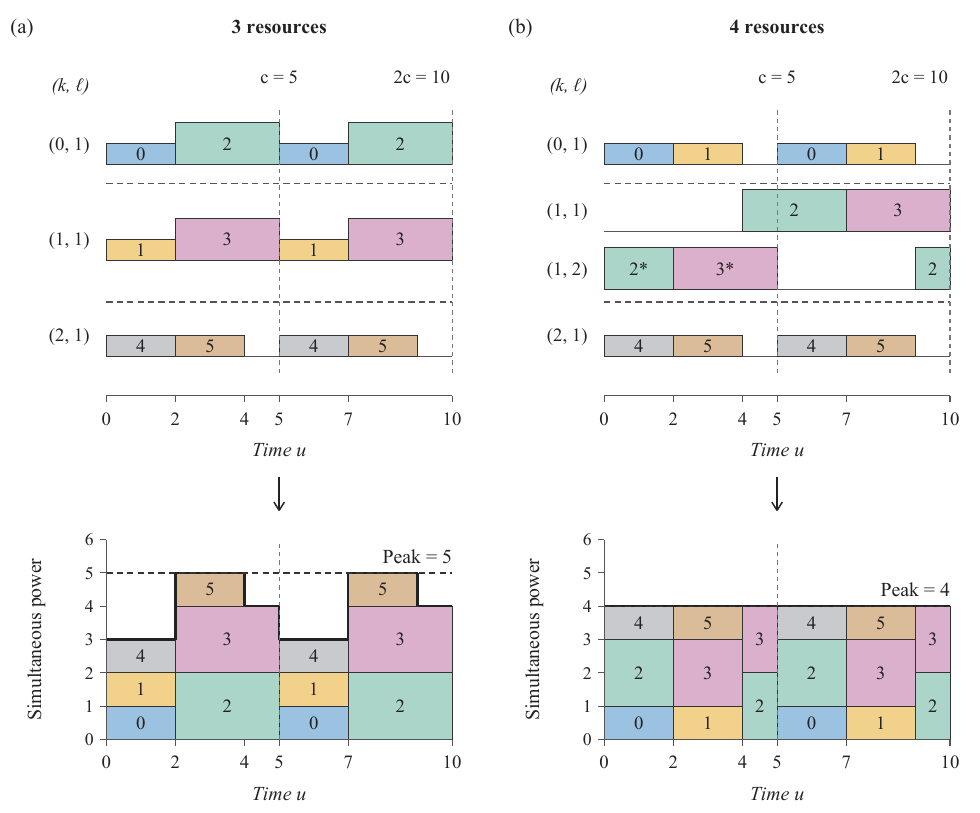}
\caption{Steady-state resource schedules and total power at $c=5$: (a) three resources, peak 5; (b) four resources, peak 4. Stars denote preceding-product copies.}
\label{fig:problem-structure}
\end{figure}

Figures~\ref{fig:line-overview} and~\ref{fig:problem-structure} illustrate these decisions on a constructed six-task example with $c=5$, $r_{\max}=2$, and $R_{\max}=4$. This example is also used to explain the SAT variables and is separate from the experimental benchmark. Figure~\ref{fig:line-overview}(a) gives task durations above the nodes and power requirements below; panel~(b) shows a resource configuration $(r_0,r_1,r_2)=(1,2,1)$ with crossover between workstations. Panel~(c) compares virtual scheduling capacity: tasks 2 and 3 each take three time units, so their combined workload of six requires a ten-unit horizon when assigned to the same workstation. With one resource per workstation, the two tasks must instead be assigned to different workstations.

Figure~\ref{fig:problem-structure} compares two feasible designs with different resource allocations. In panel~(a), task sets $\{0,2\}$, $\{1,3\}$, and $\{4,5\}$ are assigned to three workstations, each with one resource, producing the power profile $(3,3,5,5,4)$ with peak 5. In panel~(b), tasks 2 and 3 share workstation 1 and occupy $[4,7)$ and $[7,10)$ in its virtual schedule. The second resource repeats this schedule with a five-unit offset, yielding profile $(4,4,4,4,4)$ and peak 4. Although both designs consume $\sum_j t_jw_j=20$ energy units per cycle, their joint resource allocation, task assignment, and scheduling decisions lead to different power peaks. Additional resources therefore provide scheduling flexibility without guaranteeing a lower peak.

The upper panels show the activities of individual physical resources, indexed by $(k,\ell)$, while the lower panels aggregate their simultaneous power demand. Bar widths and heights indicate duration and power in arbitrary units. Because the resources process successive products, aligned rows can represent different products; stars identify preceding-product copies, and bars at the plot edges continue beyond the displayed interval.

\subsection{Reference time-indexed ILP formulation}\label{sec:ilp}

We use the time-indexed ILP formulation introduced by \citet{DelormeGianessi2022}, following its presentation as \MPSC{} in their extended study \citeyearpar{DelormeGianessi2024}. We retain its decision structure and constraints using the notation in Table~\ref{tab:notation}. The binary variables $X_{j,k}$, $S_{j,t}$, and $R_{k,r}$ represent task assignment, starting times, and resource activation, respectively.

The activity expression is
\begin{equation}
F(j,t)=
\sum_{\tau=\max\{0,t-t_j+1\}}^{\min\{t,H-t_j\}}S_{j,\tau},
\qquad j\in O,\ t\in\mathcal T.
\tag{ILP-1}\label{eq:ilp-activity}
\end{equation}
Here, $F(j,t)=1$ precisely when task $j$ is processed at virtual time $t$. Empty sums are zero.

\begin{table}[H]
\centering
\caption{Problem and ILP notation.}\label{tab:notation}
\begingroup
\small
\setlength{\tabcolsep}{5pt}
\renewcommand{\arraystretch}{0.96}
\newlength{\notationwidth}
\setlength{\notationwidth}{\dimexpr\linewidth-4\tabcolsep\relax}
\begin{tabular}{@{}p{.12\notationwidth}p{.18\notationwidth}p{.70\notationwidth}@{}}
\toprule
Category & Symbol & Definition \\
\midrule
Set & $O$ & Set of $n$ production tasks, $O=\{0,\ldots,n-1\}$.\\
Set & $M$ & Ordered set of at most $m$ workstations, $M=\{0,\ldots,m-1\}$.\\
Set & $E$ & Precedence arcs; $(i,j)\in E$ means $i\prec j$.\\
Set & $\mathcal R$ & Resource indices, $\mathcal R=\{1,\ldots,r_{\max}\}$.\\
Set & $\mathcal T$ & Virtual time slots, $\mathcal T=\{0,\ldots,H-1\}$.\\
Set & $\mathcal C$ & Physical cycle phases, $\mathcal C=\{0,\ldots,c-1\}$.\\
Set & $\mathcal T^j$ & Candidate start times of task $j$, $\{0,\ldots,H-t_j\}$.\\
Set & $\mathcal T_r^j$ & Candidate start times if at most the first $r$ resource layers are available; $\{0,\ldots,rc-t_j\}$ when nonempty.\\
Parameter & $t_j$ & Integer processing time of task $j$.\\
Parameter & $w_j$ & Integer active power demand of task $j$.\\
Parameter & $c$ & Prescribed cycle time.\\
Parameter & $r_{\max}$ & Maximum number of parallel resources at one workstation.\\
Parameter & $R_{\max}$ & Maximum number of resources over the complete line.\\
Parameter & $H$ & Augmented timespan, $H=r_{\max}c$.\\
ILP & $X_{j,k}$ & One iff task $j$ is assigned to workstation $k$.\\
ILP & $S_{j,t}$ & One iff task $j$ starts at virtual time $t$.\\
ILP & $R_{k,r}$ & One iff the $r$th resource layer of workstation $k$ is activated; monotonicity makes this equivalent to ``at least $r$ resources''.\\
ILP & $F(j,t)$ & Binary activity expression: one iff task $j$ is processed at virtual time $t$.\\
ILP & $W_M$ & Nonnegative integer upper bound on the cyclic power peak.\\
\bottomrule
\end{tabular}
\endgroup
\end{table}

The complete formulation is
\begingroup
\allowdisplaybreaks[1]
\begin{gather}
\min\ W_M
\tag{ILP-2}\label{eq:ilp-objective} \\
\text{subject to}\notag \\
\sum_{k\in M}X_{j,k}=1,
\qquad \forall j\in O,
\tag{ILP-3}\label{eq:ilp-assignment} \\
\sum_{j\in O}t_jX_{j,k}
\le c\sum_{r\in\mathcal R}R_{k,r},
\qquad \forall k\in M,
\tag{ILP-4}\label{eq:ilp-workload} \\
X_{j,k}\le \sum_{\substack{h\in M\\h\le k}}X_{i,h},
\qquad \forall (i,j)\in E,\ k\in M,
\tag{ILP-5}\label{eq:ilp-station-precedence} \\
\sum_{t\in\mathcal T^j}S_{j,t}=1,
\qquad \forall j\in O,
\tag{ILP-6}\label{eq:ilp-start} \\
X_{j,k}-R_{k,r}
\le \sum_{t\in\mathcal T_{r-1}^{j}}S_{j,t},
\qquad \forall j\in O,\ k\in M,\ r\in\mathcal R,
\tag{ILP-7}\label{eq:ilp-resource-horizon} \\
S_{j,t}\le
\sum_{\tau=0}^{t-t_i}S_{i,\tau}
+2-X_{i,k}-X_{j,k},
\quad \forall (i,j)\in E,\ k\in M,\ t\in\mathcal T^j,
\tag{ILP-8}\label{eq:ilp-time-precedence} \\
X_{i,k}+X_{j,k}+F(i,t)+F(j,t)\le 3,
\quad \forall i,j\in O:i<j,\ k\in M,\ t\in\mathcal T.
\tag{ILP-9}\label{eq:ilp-nonoverlap} \\
\sum_{j\in O}\sum_{r\in\mathcal R}
 w_jF\bigl(j,(r-1)c+\tau\bigr)
\le W_M,
\qquad \forall \tau\in\mathcal C.
\tag{ILP-10}\label{eq:ilp-peak} \\
R_{k,r+1}\le R_{k,r},
\qquad \forall k\in M,\ r=1,\ldots,r_{\max}-1,
\tag{ILP-11}\label{eq:ilp-resource-order} \\
\sum_{k\in M}\sum_{r\in\mathcal R}R_{k,r}\le R_{\max}.
\tag{ILP-12}\label{eq:ilp-resource-budget} \\
X_{j,k},S_{j,t},R_{k,r}\in\{0,1\},
\qquad W_M\in\mathbb Z_{+}.
\tag{ILP-13}\label{eq:ilp-domains}
\end{gather}
\endgroup

The assignment and starting-time constraints select one workstation and one start time per task, as specified in Eqs.~\eqref{eq:ilp-assignment} and~\eqref{eq:ilp-start}. Workstation capacity, precedence, and non-overlap are enforced by Eqs.~\eqref{eq:ilp-workload}, \eqref{eq:ilp-station-precedence}, \eqref{eq:ilp-time-precedence}, and~\eqref{eq:ilp-nonoverlap}. Resource allocation is coupled with scheduling through Eq.~\eqref{eq:ilp-resource-horizon}: if the $r$th resource is inactive, an assigned task must start in $\mathcal T_{r-1}^{j}$ and finish by $(r-1)c$. Cyclic power aggregation in Eq.~\eqref{eq:ilp-peak} accounts for activities separated by integer multiples of $c$ when bounding the peak. Finally, consecutive resource activation in Eq.~\eqref{eq:ilp-resource-order} makes $\sum_{r\in\mathcal R}R_{k,r}$ the number of resources at workstation $k$, subject to the global resource budget in Eq.~\eqref{eq:ilp-resource-budget}.

\section{SAT models for PSCALB3PM}\label{sec:sat}

We first present the Basic SAT Encoding (BSE), which adapts the SAT formulation of \citet{PyEtAl2024} to PSCALB3PM by extending the original encoding to accommodate the problem's additional constraints. We then introduce the Compact SAT Encoding (CSE), which extends the SAT encoding proposed by \cite{kieu2026compact} for SALB3PM to PSCALB3PM. Finally, we describe the encoding design for incremental SAT solving to minimize peak power consumption, which is applied to both BSE and CSE encodings.

\begingroup
\small
\setlength{\tabcolsep}{5pt}
\renewcommand{\arraystretch}{0.96}
\newlength{\satnotationwidth}
\setlength{\satnotationwidth}{\dimexpr\linewidth-2\tabcolsep\relax}
\renewcommand{\theHtable}{sat-notation.\arabic{table}}
\begin{longtable}{@{}p{.18\satnotationwidth}p{.82\satnotationwidth}@{}}
\caption{SAT notation.}\label{tab:sat-notation}\\
\toprule
Symbol & Definition \\
\midrule
\endfirsthead
\multicolumn{2}{l}{\tablename\ \thetable\ (continued)}\\
\toprule
Symbol & Definition \\
\midrule
\endhead
\bottomrule
\endfoot
$A_{j,t}$ & Activity variable: true when task $j$ is active at virtual time $t$.\\
$Y_{j,k}$ & Cumulative assignment variable: true iff task $j$ is assigned to a workstation not exceeding $k$.\\
$T_{j,t}$ & Cumulative start-time variable: true iff task $j$ starts no later than $t$.\\
$U_{\ell}$ & Peak indicator variable; $\neg U_{\ell}$ imposes peak bound $\ell$.\\
$LB,UB,B$ & Lower bound, initial upper bound, and candidate peak bound, respectively.\\
$W_{\mathrm{best}}$ & Current incumbent peak.\\
$q(j,t)$ & Minimum resource layer required by start $t$: $\lceil(t+t_j)/c\rceil$.\\
\end{longtable}
\endgroup

\subsection{Basic SAT Encoding (BSE)}

BSE adapts the serial SAT formulation of \citet{PyEtAl2024} from $c$ to the augmented timespan $H$, with start domain $\mathcal T^j$.
BSE uses four groups of Boolean variables, as summarized in Tables \ref{tab:notation} and \ref{tab:sat-notation}: ($X_{j,k}$) indicates the assignment of task ($j$) to workstation ($k$), ($S_{j,t}$) its start time, ($A_{j,t}$) its activity at time ($t$), and ($R_{k,r}$) the activation of resource layer ($r$) at workstation ($k$). 
The encoding enforces assignment and precedence, activity and non-overlap, resource and peak-power constraints, as detailed below.

\paragraph{Assignment and precedence}
Each assignment and start-time decision uses an at-least-one clause and pairwise at-most-one clauses:
\begin{align}
\bigvee_{k\in M}X_{j,k},\qquad
&\neg X_{j,k}\lor\neg X_{j,h},
&&\forall j\in O,\ k,h\in M:k<h,
\tag{SAT-D1}\label{eq:direct-assignment}\\
\bigvee_{t\in\mathcal T^j}S_{j,t},\qquad
&\neg S_{j,t}\lor\neg S_{j,u},
&&\forall j\in O,\ t,u\in\mathcal T^j:t<u.
\tag{SAT-D2}\label{eq:direct-start}
\end{align}
Different-station precedence is enforced by SAT-D3, which excludes assignment pairs that place a successor task at an earlier workstation than its predecessor:
\begin{equation}
\neg X_{i,h}\lor\neg X_{j,k},
\qquad
\forall(i,j)\in E,\ h,k\in M:h>k,
\tag{SAT-D3}\label{eq:direct-station-precedence}
\end{equation}
Same-station precedence is enforced by SAT-D4, which excludes start-time pairs that violate the temporal order of two tasks assigned to the same workstation:
\begin{equation}
\neg X_{i,k}\lor\neg X_{j,k}\lor
\neg S_{i,t}\lor\neg S_{j,u},
\quad
\begin{aligned}
&\forall(i,j)\in E,\ k\in M,\\[-1mm]
&t\in\mathcal T^i,\ u\in\mathcal T^j:t+t_i>u.
\end{aligned}
\tag{SAT-D4}\label{eq:direct-time-precedence}
\end{equation}

\paragraph{Activity and non-overlap}
The activity variables represent the processing intervals induced by the selected start times. Specifically, if task \(j\) starts at time \(t\), it must be active at every time point in its processing interval:
\begin{equation}
\begin{aligned}
&S_{j,t}\Longrightarrow A_{j,t+\epsilon},\\
&\qquad \forall j\in O,\ t\in\mathcal T^j,\ \epsilon=0,\ldots,t_j-1.
\end{aligned}
\tag{SAT-1}\label{eq:sat-activity}
\end{equation}

Using these activity variables, SAT-2 prevents two tasks assigned to the same workstation from being processed simultaneously:
\begin{equation}
\left(X_{i,k}\land X_{j,k}\right)
\Longrightarrow
\left(\neg A_{i,t}\lor\neg A_{j,t}\right),
\quad
\forall i<j,\ k\in M,\ t\in\mathcal T.
\tag{SAT-2}\label{eq:sat-nonoverlap}
\end{equation}

\paragraph{Resource and peak constraints}
The resource constraints ensure that each task can be scheduled within the available resource layers while enforcing consecutive activation of resource layers at each
workstation:
\begin{equation}
R_{k,r+1}\Longrightarrow R_{k,r},
\qquad
\forall k\in M,\ r=1,\ldots,r_{\max}-1.
\tag{SAT-3}\label{eq:sat-resource-order}
\end{equation}
If task \(j\) starts at time \(t\) and is assigned to workstation \(k\), the
required resource layer is
\(q(j,t)=\lceil(t+t_j)/c\rceil\), which must be activated at that workstation:
\begin{equation}
X_{j,k}\land S_{j,t}\Longrightarrow R_{k,q(j,t)},
\quad
\forall j\in O,\ k\in M,\ t\in\mathcal T^j.
\tag{SAT-4}\label{eq:sat-resource-link}
\end{equation}
Since \(t+t_j\le H=r_{\max}c\), the required layer satisfies
\(1\le q(j,t)\le r_{\max}\). The total number of activated resource layers
across the line is bounded by \(R_{\max}\):
\begin{equation}
\sum_{k\in M}\sum_{r\in\mathcal R}R_{k,r}\le R_{\max}.
\tag{SAT-5}\label{eq:sat-resource-budget}
\end{equation}

To constrain peak power, the augmented timespan is mapped to the physical
cycle phases by grouping time points that are congruent modulo \(c\).
For a candidate peak bound \(B\), the total power demand in each physical
phase must not exceed \(B\):
\begin{equation}
\sum_{j\in O}w_j\sum_{q=0}^{r_{\max}-1}A_{j,\tau+qc}
\le B,
\qquad \forall\tau\in\mathcal C.
\tag{SAT-6}\label{eq:sat-peak-bound}
\end{equation}

\subsection{Compact SAT Encoding (CSE)}

CSE adapts the serial staircase encoding of \citet{kieu2026compact} to the augmented timespan $H$. In addition to BSE's variables, it introduces two cumulative variables, ($Y_{j,k}$) and ($T_{j,t}$), for workstation assignment and start-time representation, respectively. It retains BSE's activity, non-overlap, resource, and cyclic-peak constraints (SAT-1 -- SAT-6), replacing SAT-D1 -- SAT-D4 with the cumulative assignment, start-time, and precedence constraints SAT-7 -- SAT-16 below.

\paragraph{Cumulative variables}

For $(i,j)\in E$, assigning task $i$ to workstation $k+1$ excludes $j$ from
workstations $0,\ldots,k$:
\[
X_{i,k+1}+\sum_{h=0}^{k}X_{j,h}\le 1,
\qquad k=0,\ldots,m-2.
\]
Different-station and same-station precedence share prefixes of assignment and
start-time decisions, represented by the cumulative variables
\begin{equation*}
Y_{j,k}\ \Longleftrightarrow\ \bigvee_{h=0}^{k}X_{j,h},
\qquad
T_{j,t}\ \Longleftrightarrow\ \bigvee_{\tau=0}^{t}S_{j,\tau}.
\end{equation*}
Only $Y_{j,k}$ for $k=0,\ldots,m-2$ and $T_{j,t}$ for $t=0,\ldots,L_j-1$
are created: the final cumulative values are true under exactly-one
assignment and start decisions.

\begin{figure}[htbp]
\centering
\includegraphics[width=.98\linewidth]{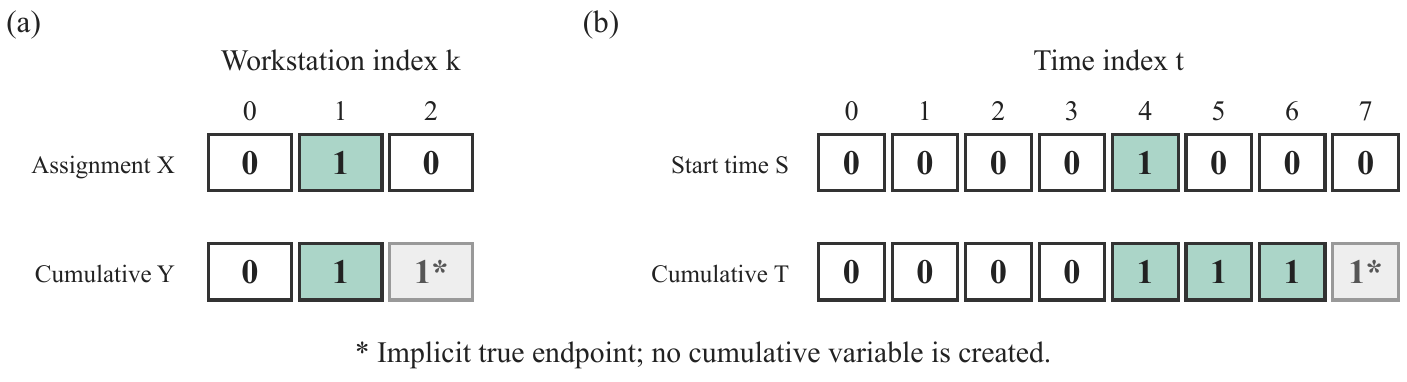}
\caption{Base and cumulative values for task 2: (a) workstation assignment; (b) start time. Shading denotes true entries; starred endpoints are implicit constants.}
\label{fig:encoding-example}
\end{figure}

Figure~\ref{fig:encoding-example} illustrates task 2 in schedule B, assigned to
workstation 1 with $s_2=4$, $H=10$, $t_2=3$, and $L_2=7$. In panels~(a) and~(b),
the cumulative values switch from false to true at the selected workstation and
start time. CSE retains $X$ and $S$ and reduces pairwise clauses by sharing these
prefixes across precedence constraints.

\paragraph{Constraint formulation}
The cumulative workstation-assignment encoding is
\begin{align}
Y_{j,0}&\Longleftrightarrow X_{j,0},
&&\forall j\in O,
\tag{SAT-7}\label{eq:sat-x-base}\\
Y_{j,k-1}&\Longrightarrow Y_{j,k},
&&\forall j\in O,\ k=1,\ldots,m-2,
\tag{SAT-8}\label{eq:sat-x-monotone}\\
X_{j,k}&\Longleftrightarrow
\bigl(Y_{j,k}\land\neg Y_{j,k-1}\bigr),
&&\forall j\in O,\ k=1,\ldots,m-2,
\tag{SAT-9}\label{eq:sat-x-step}\\
X_{j,m-1}&\Longleftrightarrow \neg Y_{j,m-2},
&&\forall j\in O.
\tag{SAT-10}\label{eq:sat-x-last}
\end{align}
Equations~\eqref{eq:sat-x-base}--\eqref{eq:sat-x-last} replace SAT-D1 and select
exactly one workstation. Standard translation into conjunctive normal form (CNF) gives three clauses for each step
equivalence and one for each monotonicity implication. Different-station precedence
requires one binary clause per threshold:
\begin{equation}
X_{i,k+1}\Longrightarrow\neg Y_{j,k},
\qquad
\forall(i,j)\in E,\ k=0,\ldots,m-2.
\tag{SAT-11}\label{eq:sat-station-precedence}
\end{equation}
This replaces SAT-D3 with $O(|E|m)$ clauses.

The analogous cumulative start-time constraints replace SAT-D2:
\begin{equation}
\begin{cases}
T_{j,0}\Longleftrightarrow S_{j,0},&L_j\ge1,\\
S_{j,0},&L_j=0,
\end{cases}
\qquad \forall j\in O,
\tag{SAT-12}\label{eq:sat-t-base}
\end{equation}
\begin{align}
T_{j,t-1}&\Longrightarrow T_{j,t},
&&\forall j\in O,\ t=1,\ldots,L_j-1,
\tag{SAT-13}\label{eq:sat-t-monotone}\\
S_{j,t}&\Longleftrightarrow
\bigl(T_{j,t}\land\neg T_{j,t-1}\bigr),
&&\forall j\in O,\ t=1,\ldots,L_j-1,
\tag{SAT-14}\label{eq:sat-t-step}\\
S_{j,L_j}&\Longleftrightarrow\neg T_{j,L_j-1},
&&\forall j\in O:L_j\ge1.
\tag{SAT-15}\label{eq:sat-t-last}
\end{align}
The final cumulative value $T_{j,L_j}$ is implicit.

To replace the start pairs in SAT-D4, let $\theta(i,t)=t+t_i-1$ be the latest
forbidden start for successor $j$ when $i$ starts at $t$:
\begin{equation}
\left(X_{i,k}\land X_{j,k}\land S_{i,t}\right)
\Longrightarrow
\begin{cases}
\neg T_{j,\theta(i,t)},&\theta(i,t)<L_j,\\
\bot,&\theta(i,t)\ge L_j,
\end{cases}
\quad
\begin{aligned}
&\forall(i,j)\in E,\\[-1mm]
&k\in M,\ t\in\mathcal T^i.
\end{aligned}
\tag{SAT-16}\label{eq:sat-time-precedence}
\end{equation}
The branches enforce completion before $j$ starts or exclude the placement when
no successor start remains. Equivalent head, middle, and tail CNF ranges avoid
out-of-domain cumulative variables.

\paragraph{Exactness of the feasibility encoding}
For the discrete-time problem defined in Section~\ref{sec:problem}, with $t_j\le H$ for every task, SAT-1--SAT-16
(with the boundary conventions above) is satisfiable at bound $B$ if and only if
a feasible schedule with power peak at most $B$ exists.

Given a feasible schedule, set $X$ and $S$ to its decisions, $Y$ and $T$ to their
prefixes, and $A$ to the processing intervals. These satisfy SAT-7--SAT-16
and the shared constraints SAT-1--SAT-2.
Activating the first $r_k$ resource layers satisfies SAT-3--SAT-5, since
completion within $r_kc$ implies $q(j,t)\le r_k$ and the schedule respects the
resource budget. Its phase demand satisfies SAT-6.

Conversely, SAT-7--SAT-15 select exactly one workstation and one start per task
and preserve workstation order; SAT-16 enforces same-station precedence. SAT-1
covers every processing interval with true activity literals, so SAT-2 prevents overlap.
With $r_k=\sum_{r\in\mathcal R}R_{k,r}$, SAT-3--SAT-4 place these disjoint
intervals within $[0,r_kc)$. Their total length is therefore at most $r_kc$,
establishing the ILP workload inequality. SAT-5 enforces the resource budget.
Since $w_j>0$, actual phase demand is bounded by the activity sum in SAT-6 and
hence by $B$. Unsupported true activity literals only strengthen the non-overlap
and peak restrictions. Equisatisfiable CNF translations preserve this equivalence.

\paragraph{Encoding size}

Table~\ref{tab:sat-size} gives the encoding size over the augmented timespan $H$ before
weighted constraints are translated into CNF. The cumulative bounds follow
\citet{kieu2026compact}.

\begin{table}[htbp]
\centering
\caption{CSE encoding size.}
\label{tab:sat-size}
\small
\begin{tabular}{@{}p{0.43\linewidth}p{0.21\linewidth}p{0.25\linewidth}@{}}
\toprule
Component & Variable order & Clauses or weighted input terms\\
\midrule
Cumulative assignment, SAT-7--SAT-10 & $O(nm)$ & $O(nm)$\\
Cumulative start time, SAT-12--SAT-15 & $O(nH)$ & $O(nH)$\\
Different-station precedence, SAT-11 & -- & $O(|E|m)$\\
Same-station precedence, SAT-16 & -- & $O(|E|mH)$\\
Activity propagation, SAT-1 & $O(nH)$ & $O\!\left(\sum_j|\mathcal T^j|t_j\right)$\\
Non-overlap, SAT-2 & -- & $O(n^2mH)$\\
Resource constraints, SAT-3--SAT-5 & $O(mr_{\max})$ & $O(nmH+mr_{\max})$;\newline one CNF encoding\\
One phase-bound family, SAT-6 & encoder-dependent & $O(nH)$ weighted literals\\
\bottomrule
\end{tabular}
\end{table}
CSE removes quadratic workstation and temporal-precedence comparisons, while the
$O(n^2mH)$ non-overlap term remains unchanged. Resource-budget and peak constraints
use the binary-merger encoding of \citet{MantheyEtAl2014} in the PySAT toolkit
\citep{IgnatievEtAl2018}, with CaDiCaL as the solver \citep{PollittEtAl2026}.

\subsection{Incremental peak optimization}\label{sec:incremental}

Both encodings use the ordered peak-indicator strategy of
\citet{kieu2026compact}, adapted to resource-dependent time horizons and cyclic phase
demand. The solution-dependent activity no-goods of \citet{PyEtAl2024} are
omitted. Algorithm~\ref{alg:incremental-peak} summarizes the procedure.

Let $\Phi_{\mathrm{base}}$ contain the shared constraints SAT-1--SAT-5 and
either SAT-D1--SAT-D4 for BSE or SAT-7--SAT-16 for CSE, without a fixed peak bound.
A single solve supplies the initial schedule and bounds:
\begin{equation}
LB=\max_{j\in O}w_j,
\qquad
UB=\max_{\tau\in\mathcal C}P_\tau,
\tag{SAT-17}\label{eq:sat-initial-bounds}
\end{equation}
Every task is active during at least one phase, giving $LB$. For $UB$ and all
later incumbents, \textsc{Peak} reconstructs phase demand from the selected starts
using Eq.~\eqref{eq:problem-phase-power}, excluding unsupported activity literals.
This single-solve initialization replaces the multiple bound-initialization
solves of \citet{kieu2026compact}.

For the fixed initial interval $[LB,UB]$, introduce peak indicator variables
$U_\ell$, $\ell=LB+1,\ldots,UB-1$, ordered by
\begin{equation}
\neg U_{\ell}\lor U_{\ell-1},
\qquad \ell=LB+2,\ldots,UB-1.
\tag{SAT-18}\label{eq:sat-bound-ladder}
\end{equation}
For every phase $\tau$, one binary-merger CNF encoding is generated for
\begin{equation}
\sum_{j\in O}w_j\sum_{q=0}^{r_{\max}-1}A_{j,\tau+qc}
+\sum_{\ell=LB+1}^{UB-1}(1-U_{\ell})
\le UB.
\tag{SAT-19}\label{eq:sat-augmented-peak}
\end{equation}
A candidate bound $B\in\{LB+1,\ldots,UB-1\}$ is activated by the unit clause
\begin{equation}
\neg U_B.
\tag{SAT-20}\label{eq:sat-activate-bound}
\end{equation}
Ordering forces $U_\ell$ false for $\ell\ge B$, adding at least $UB-B$ penalty
terms in Eq.~\eqref{eq:sat-augmented-peak} and limiting phase demand to $B$.
Conversely, any schedule meeting $B$ extends to a satisfying assignment by setting
$U_\ell$ true for $\ell<B$ and false otherwise, giving exactly $UB-B$ penalty terms.

Starting at $UB$, each satisfiable call updates the incumbent to reconstructed
peak $W_{\mathrm{best}}$. If $W_{\mathrm{best}}=LB$, it is optimal; otherwise, the next test uses $B=W_{\mathrm{best}}-1$.
For $B>LB$, adding SAT-20 retains the formula, solver state, and learned clauses.
At $B=LB$, SAT-6 is encoded directly because $U_{LB}$ is not defined.
An unsatisfiable strict-bound test certifies optimality.

In Algorithm~\ref{alg:incremental-peak}, \textsc{Timeout} denotes interruption
by the experiment controller at the SAT process limit, during construction or
search. The latest saved incumbent, if available, is retained without an
optimality claim; its timestamp records the incumbent update, not search completion.

\begin{algorithm}[htbp]
\caption{Shared incremental peak optimization for \PSCALB{}}
\label{alg:incremental-peak}
\begin{singlespace}
\small
\begin{algorithmic}[1]
\Require Feasibility formula $\Phi_{\mathrm{base}}$, SAT process time limit
\Ensure Best feasible schedule and its certification status, if available
\State Initialize solver $\mathcal S$ with $\Phi_{\mathrm{base}}$
\State $\phi\gets\Call{Solve}{\mathcal S}$
\State \textbf{if} $\phi=\textsc{Unsat}$ \textbf{then return} \textsc{Infeasible}
\State \textbf{if} $\phi=\textsc{Timeout}$ \textbf{then return} no incumbent, \textsc{Timeout}
\State $\phi_{\mathrm{best}}\gets\phi$; $UB\gets\Call{Peak}{\phi}$; $W_{\mathrm{best}}\gets UB$; $LB\gets\max_{j\in O}w_j$
\State Add $U_\ell$, $\ell=LB+1,\ldots,UB-1$, and ordering clauses SAT-18 to $\mathcal S$
\For{each physical phase $\tau\in\mathcal C$}
    \State Add the binary-merger CNF encoding of SAT-19 to $\mathcal S$
\EndFor
\While{SAT process budget remains}
    \State $\phi\gets\Call{Solve}{\mathcal S}$
    \State \textbf{if} $\phi=\textsc{Timeout}$ \textbf{then break}
    \State \textbf{if} $\phi=\textsc{Unsat}$ \textbf{then return} $(\phi_{\mathrm{best}},W_{\mathrm{best}},\textsc{Optimal})$
    \State $\phi_{\mathrm{best}}\gets\phi$; $W_{\mathrm{best}}\gets\Call{Peak}{\phi}$
    \State \textbf{if} $W_{\mathrm{best}}=LB$ \textbf{then return} $(\phi_{\mathrm{best}},W_{\mathrm{best}},\textsc{Optimal})$
    \If{$W_{\mathrm{best}}-1=LB$}
        \State Add the CNF encoding of SAT-6 with $B=LB$ to $\mathcal S$
    \Else
        \State Add the unit clause $\neg U_{W_{\mathrm{best}}-1}$ to $\mathcal S$
    \EndIf
\EndWhile
\State \Return $(\phi_{\mathrm{best}},W_{\mathrm{best}},\textsc{Feasible})$
\end{algorithmic}
\end{singlespace}
\end{algorithm}

\section{Computational study}\label{sec:experiments}

\subsection{Research questions}

The experiments are organized around the following questions:
\begin{description}[style=nextline,leftmargin=2.5cm]
    \item[RQ1.] \begin{minipage}[t]{\linewidth}How do BSE and CSE compare with \MPSC{} and \MSLS{} in solution quality and feasibility, and with \MPSC{} in optimality coverage and proof-oriented charged time?\end{minipage}
    \item[RQ2.] \begin{minipage}[t]{\linewidth}Under the same incremental optimizer, what changes in solution quality, optimality coverage, charged time, and base-CNF size result from replacing BSE with CSE?\end{minipage}
\end{description}

\subsection{Experimental setup}\label{sec:experimental-setup}

The benchmark set is exactly the one introduced by \citet{DelormeGianessi2024}: 17 base \SALB{} instances with up to 30 tasks, 14 workstations, and cycle times up to 94. For a base instance with $m$ workstations, the admissible values of $R_{\max}$ yield one instance for $r_{\max}=1$, $m+1$ instances for $r_{\max}=2$, and $2m+1$ instances for $r_{\max}=3$. The study therefore evaluates 17, 125, and 233 instances in the three respective resource groups, for exactly 375 \PSCALB{} instances in total. 
The four-method assessment includes two SAT representations and two reference methods:
\begin{enumerate}[label=(\alph*)]
    \item \emph{Basic SAT Encoding (BSE)}, which extends the Boolean workstation-assignment and
    start-time representation of \citet{PyEtAl2024} to the augmented timespan $H$;
    \item \emph{Compact SAT Encoding (CSE)}, the proposed method using the cumulative
    representation in SAT-1--SAT-16;
    \item the time-indexed ILP model \MPSC{} of \citet{DelormeGianessi2024}; and
    \item the \MSLS{} matheuristic of \citet{DelormeGianessi2024}.
\end{enumerate}
Both SAT variants use the incremental peak optimization of
Subsection~\ref{sec:incremental}. The names BSE and CSE identify
their Boolean representations under this common optimizer. RQ1 compares both SAT variants with the
reference methods, while RQ2 isolates their representation difference.
The \MSLS{} implementation follows the search and decoding procedure of \citet{DelormeGianessi2024}, to which we refer for the solution coding, neighborhood, and decoder. Its experimental budget and replication protocol are specified below. All four methods are evaluated on the common benchmark and use the same comparison reference in the results.

The experiments were executed on a Google Cloud Platform \texttt{c2d-highmem-8} virtual machine with an AMD EPYC~7B13 processor, 8 vCPUs, and 64~GB of provisioned memory (approximately 62~GiB reported by the operating system). The machine ran Debian GNU/Linux~13 (trixie), with Linux kernel \texttt{6.12.101+deb13-cloud-amd64}, and Python~3.13.5. The SAT implementations use PySAT (\texttt{python-sat}~1.9.dev15) \citep{IgnatievEtAl2018} with CaDiCaL~3.0.0 \citep{PollittEtAl2026} through the \texttt{Cadical300} interface. The reference methods use IBM ILOG CPLEX Callable Library~22.2.0.0, accessed through the \texttt{cplex} Python package~22.2.0.1 and DOcplex~2.32.264.
The runs use a nominal per-instance limit of $T=3600$~s. Reported times follow the common charged-time convention in Eq.~\eqref{eq:eval-time}. The implementation supports $r_{\max}\in\{1,2,3\}$, the complete per-workstation resource range of the benchmark; larger resource limits are outside the computational scope of this study.

For each solution coding, the \MSLS{} decoder uses a 60~s limit for its initial exact scheduling evaluation at the coding's base resource allocation and a 10~s limit for each trial that adds a resource, subject to the remaining search budget. Local search stops after 50 consecutive non-improving steps, after which the multi-start procedure can begin another search while time remains. Unlike the 600~s, 20-replication protocol of \citet{DelormeGianessi2024}, the present comparison uses the nominal 3600~s budget and one recorded run per instance.

For the ablation, both SAT variants use identical feasible-domain preprocessing,
activity and resource variables, resource-budget and cyclic-peak encodings, CaDiCaL
configuration, lower-bound rule, strict-improvement policy, stopping conditions, and
time accounting. The only controlled change
between the SAT variants is whether workstation and start-time
decisions and their precedence relations use the direct clauses SAT-D1--SAT-D4 or
the cumulative clauses SAT-7--SAT-16. This design holds the optimization procedure fixed while changing the representation.
Both reference implementations explicitly set the CPLEX thread count to one, including each internal decoder call in \MSLS{}. Neither overrides the CPLEX parallel-mode or random-seed parameter; these retain their library defaults. No explicit CaDiCaL seed or preprocessing override is set, so both SAT variants use the same default solver configuration. The \MSLS{} data contain one run per instance, with replication index zero and a recorded instance-specific seed. This search seed is distinct from the internal CPLEX random-seed parameter. Each reported peak is the final incumbent of that run; no aggregation across random seeds is performed.

\subsection{Performance measures}

Let $I$ be the common set of 375 \PSCALB{} instances. The method set is
\[
\mathcal A=\{\mathrm{BSE},\MPSC,\MSLS,\mathrm{CSE}\}.
\]
Both SAT representations use the same incremental optimizer. A base \SALB{} instance is
identified by $f=(\mathrm{name},n,m,c)$, and a \PSCALB{} instance by
$i=(f,r,R)$, where $r=r_{\max}$ and $R=R_{\max}$. Thus two base instances with
the same name but different $n$, $m$, or $c$ have separate comparison values.
Let $W_{a,i}$ be the final feasible peak recorded for method $a$ on instance
$i$. The direct comparison value for that instance is
\begin{equation}
BK_i^{\mathrm{raw}}=\min_{a\in\mathcal A:\,W_{a,i}\ \mathrm{available}}W_{a,i}.
\tag{E-1}\label{eq:eval-bk-raw}
\end{equation}
An empty minimum is undefined.

Both resource parameters are upper limits. A schedule feasible at smaller
limits remains feasible when either limit increases, while $f$ is held fixed.
We therefore retain available comparison schedules through the two predecessors
$i_r^-=(f,r-1,R)$ and $i_R^-=(f,r,R-1)$:
\begin{equation}
BK_i=\min\{BK_i^{\mathrm{raw}},\ BK_{i_r^-},\ BK_{i_R^-}\}.
\tag{E-2}\label{eq:eval-bk}
\end{equation}
Only predecessors belonging to the benchmark domain of the same base instance enter
this minimum; unavailable values are omitted. Processing increasing $r$ and,
within each $r$, increasing $R$ propagates the available peaks transitively.
Equivalently, $BK_i$ is the smallest recorded feasible peak over all methods and
all benchmark instances $(f,q,Q)$ with $q\le r$ and $Q\le R$.
This inherited best-known value is a feasible upper bound for $i$.

For a feasible run, the gap is
\begin{equation}
\%BK_{a,i}=100\,\frac{W_{a,i}-BK_i}{BK_i}.
\tag{E-3}\label{eq:eval-pct-bk}
\end{equation}
The recorded method-specific peaks are used in the numerator; inheritance
updates the common reference. Consequently, every method can have a positive
gap on an instance if a predecessor supplied a better feasible schedule.
The same reference is used for the SAT ablation and the reference-method
comparison.

Let $F_a\subseteq I$ be the instances on which method $a$ returns a
feasible solution, and let $\mathcal G\subseteq I$ be a nonempty reporting
group. The mean gap is
\[
\overline{\%BK}_{a,\mathcal G}
=\frac{1}{|F_a\cap\mathcal G|}
\sum_{i\in F_a\cap\mathcal G}\%BK_{a,i}.
\]
It is undefined, and reported as n/a, if $F_a\cap\mathcal G=\varnothing$.
Because feasible subsets can differ, this conditional gap is interpreted
together with feasibility coverage.

Let $O_a\subseteq F_a$ denote the runs reporting solver-proven global
optimality. Coverage uses all instances in the reporting group:
\begin{equation}
\mathrm{feas}_{a,\mathcal G}(\%)=100\frac{|F_a\cap\mathcal G|}{|\mathcal G|},
\qquad
\mathrm{opt}_{a,\mathcal G}(\%)=100\frac{|O_a\cap\mathcal G|}{|\mathcal G|}.
\tag{E-4}\label{eq:eval-coverage}
\end{equation}
Optimality coverage is reported for the three exact methods; \MSLS{} cells contain
-- (not applicable). Matching $BK_i$ does not establish optimality. Neither
feasibility nor certification status is inherited between instances.
Certification refers to the solver's status, rather than an independently
checked proof artifact.

We report proof-oriented charged time based on each run's recorded runtime.
For time accounting, $O_{\MSLS}=\varnothing$.
With $T=3600$~s and recorded time $t_{a,i}$,
\begin{equation}
\widetilde t_{a,i}=\begin{cases}
\min\{t_{a,i},T\},&i\in O_a,\\
T,&i\notin O_a,
\end{cases}
\qquad
\overline t_{a,\mathcal G}
=\frac{1}{|\mathcal G|}\sum_{i\in\mathcal G}\widetilde t_{a,i}.
\tag{E-5}\label{eq:eval-time}
\end{equation}
Every \MSLS{} run is charged $T$ under this convention; the charge does not measure
heuristic speed. The raw peaks, statuses, and timer values remain those of the
individual runs. For each base instance, Total pools its $3m+3$ \PSCALB{} instances. Overall
uses all 375 instances for coverage and time, and each method's feasible
subset for its mean gap, rather than an unweighted mean of base-instance summaries.

\FloatBarrier
\section{Performance evaluation}\label{sec:results}

This section evaluates the proposed SAT representations with respect to the two research questions. RQ1 is addressed in Sections~\ref{sec:reference-quality} and~\ref{sec:proof-time}, which evaluate solution quality, feasibility coverage, and optimality certification against the reference methods. RQ2 is investigated in Section~\ref{sec:representation-ablation}, which compares BSE and CSE under the same incremental optimization strategy to isolate the effect of the encoding representation. Detailed experimental results aggregated by base
instance are provided in Appendix~\ref{app:comparison}. Appendix~\ref{app:encoding-summary} summarizes encoding size and the breakdown of the charged-time reduction.

\subsection{Solution quality and feasibility against reference methods}
\label{sec:reference-quality}

Both BSE and CSE find feasible solutions on all 375 instances, compared with 346 (92.27\%) for \MPSC{} and 367 (97.87\%) for \MSLS{}. The mean inherited-reference gaps are 0.144\% for BSE and 0.089\% for CSE, compared with 3.390\% for \MPSC{} and 0.786\% for \MSLS{}. These means use 375, 375, 346, and 367 feasible runs, respectively, and must be interpreted together with coverage. Table~\ref{tab:method-comparison} in Appendix~\ref{app:comparison} gives the results aggregated by base instance.

\begin{figure}[!htbp]
\centering
\includegraphics[width=\linewidth]{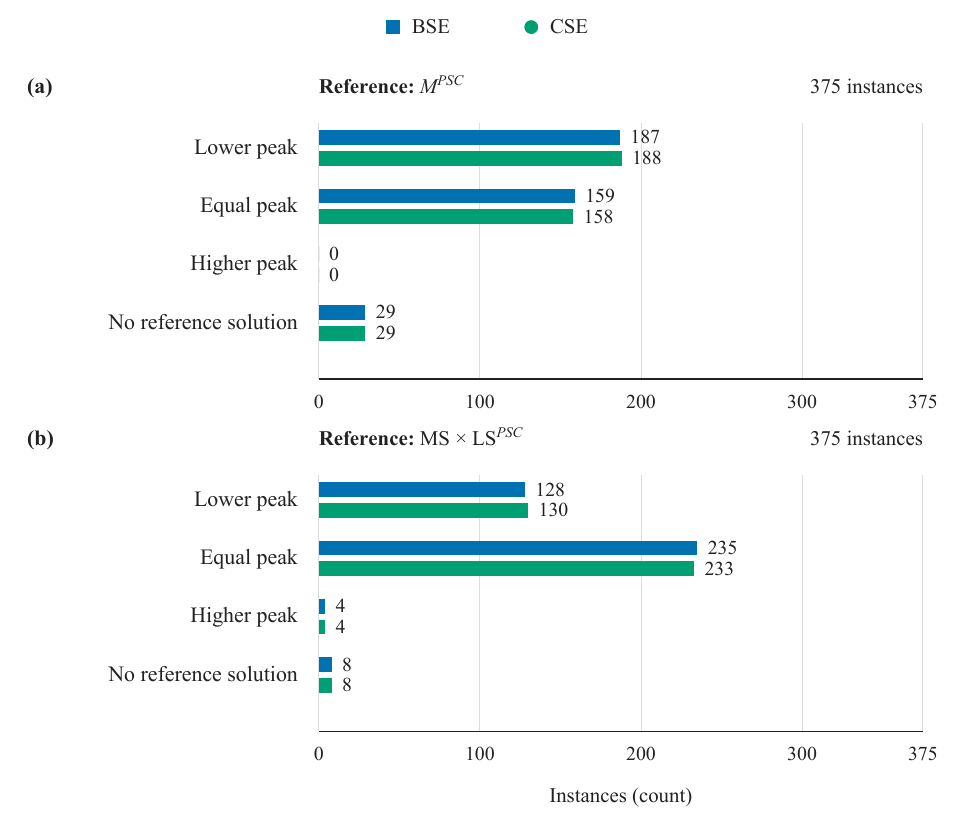}
\caption{Peak comparisons against (a) \MPSC{} and (b) \MSLS{}.}
\label{fig:reference-pairing}
\end{figure}

Figure~\ref{fig:reference-pairing}(a) compares returned peaks on the same instance, independently of BK inheritance. Against \MPSC{}, BSE returns a lower peak on 187 instances and the same peak on 159; CSE returns a lower peak on 188 and the same peak on 158. Neither SAT variant yields a higher peak, and both find feasible schedules for all $29$ instances for which the ILP model fails to find a feasible solution. Against \MSLS{}, BSE returns a lower peak on 128 instances, the same peak on 235, and a higher peak on four; the corresponding CSE counts are 130, 233, and four. Both SAT variants supply feasible schedules on the eight instances where \MSLS{} has no incumbent. Each method's four outcome counts sum to 375. Missing reference incumbents form a separate category rather than a lower-peak comparison.
Against \MSLS{}, BSE obtains a lower peak on 128 instances, matches its peak on 235, and yields a higher peak on four; the corresponding counts for CSE are 130, 233, and four. Both SAT variants find feasible schedules for all eight instances for which \MSLS{} fails to find a feasible solution (Figure~\ref{fig:reference-pairing}(b)). For each SAT variant, these three outcome counts, together with the eight instances lacking a \MSLS{} incumbent, sum to 375. Instances without a reference incumbent are treated separately rather than included in the peak comparison.

\begin{figure}[!htbp]
\centering
\includegraphics[width=\linewidth]{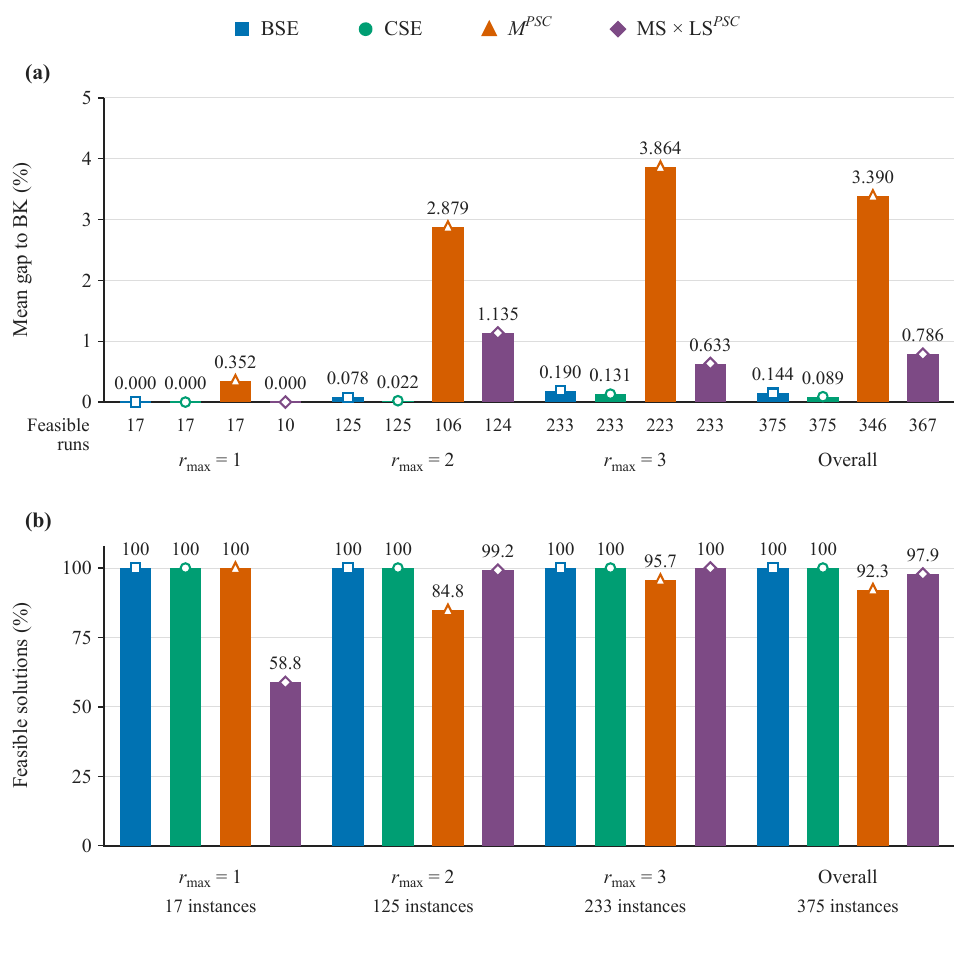}
\caption{Solution quality and feasibility: (a) mean gap to BK; (b) feasible-solution coverage.}
\label{fig:quality-feasibility}
\end{figure}

Figure~\ref{fig:quality-feasibility} summarizes the mean gap to BK and feasibility coverage by per-workstation resource limit and overall. Each instance has equal weight within its reporting group. Overall pools all 375 instances. Panel~(a) averages only feasible runs, with their counts shown below the bars, whereas panel~(b) uses every instance in the group. The common method order and colors link the two panels. In particular, the zero mean gap of \MSLS{} at $r_{\max}=1$ covers only 10 of 17 instances, so gap and coverage must be read together.

\FloatBarrier
\subsection{Optimality Certification across Time Limits}\label{sec:proof-time}

\begin{figure}[!htbp]
\centering
\includegraphics[width=\linewidth]{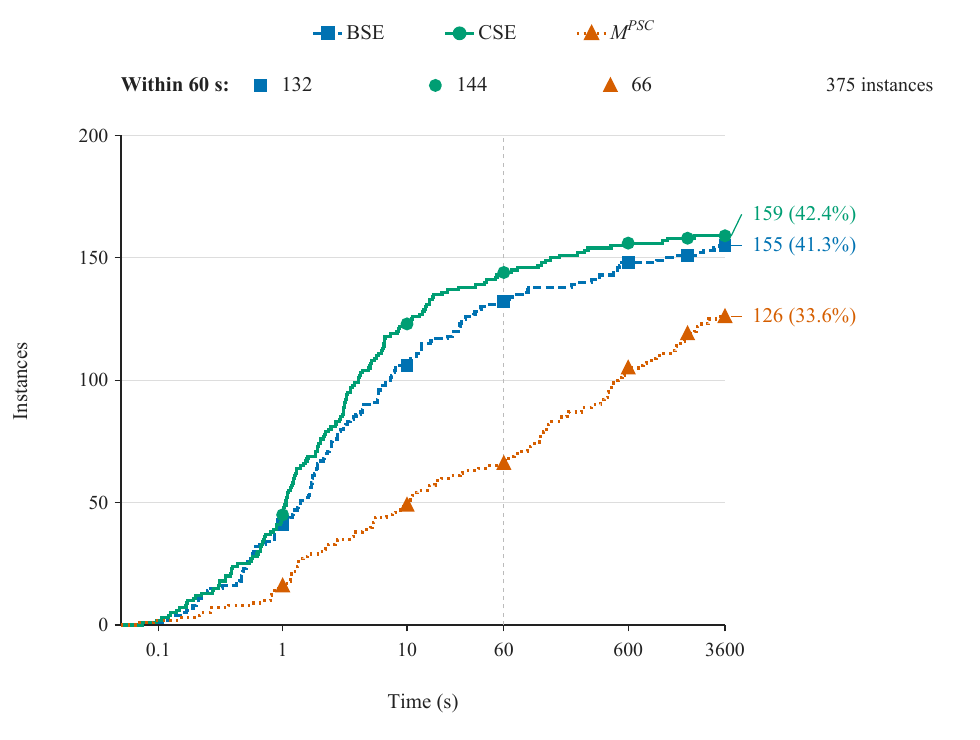}
\caption{Instances solved to optimality over time.}
\label{fig:proof-profile}
\end{figure}

BSE certifies 155 instances (41.33\%), CSE 159 (42.40\%), and \MPSC{} 126 (33.60\%). Figure~\ref{fig:proof-profile} shows cumulative optimality certificates across time thresholds. Within 60~s, BSE and CSE certify 132 and 144 instances, respectively, compared with 66 for \MPSC{}; at 600~s, the counts are 148, 156, and 105. Thus, both SAT encodings outperform the reference ILP in optimality coverage at these thresholds.
The figure includes only runs proving optimality within 3600~s, using recorded runtimes. The time axis is logarithmic, and endpoint percentages are calculated over all 375 instances. \MSLS{} is omitted because it does not prove global optimality. Appendix~\ref{app:comparison}, Table~\ref{tab:proof-time}, reports all selected thresholds.

Mean charged times are 2164.6~s for BSE, 2094.4~s for CSE, and 2513.2~s for \MPSC{}, assigning 3600~s to each uncertified run (Table \ref{tab:method-comparison} in Appendix \ref{app:comparison}). These values measure optimality certification, not time to the first feasible solution. Every instance certified by BSE or \MPSC{} is also certified by CSE within 3600~s, although individual proof times may differ. CSE has lower charged time than \MPSC{} on 153 instances, higher time on six, and equal time on 216; its mean is also lower for each of the 17 base instances.
These results are descriptive comparisons on a fixed benchmark. Since instances derived from the same base instance are related, the 375 observations are not treated as independent statistical replications. Base-instance aggregates complement the instance-weighted results.

\FloatBarrier
\subsection{Impact of the Encoding Representation: BSE versus CSE}
\label{sec:representation-ablation}

\begin{figure}[!htbp]
\centering
\includegraphics[width=\linewidth]{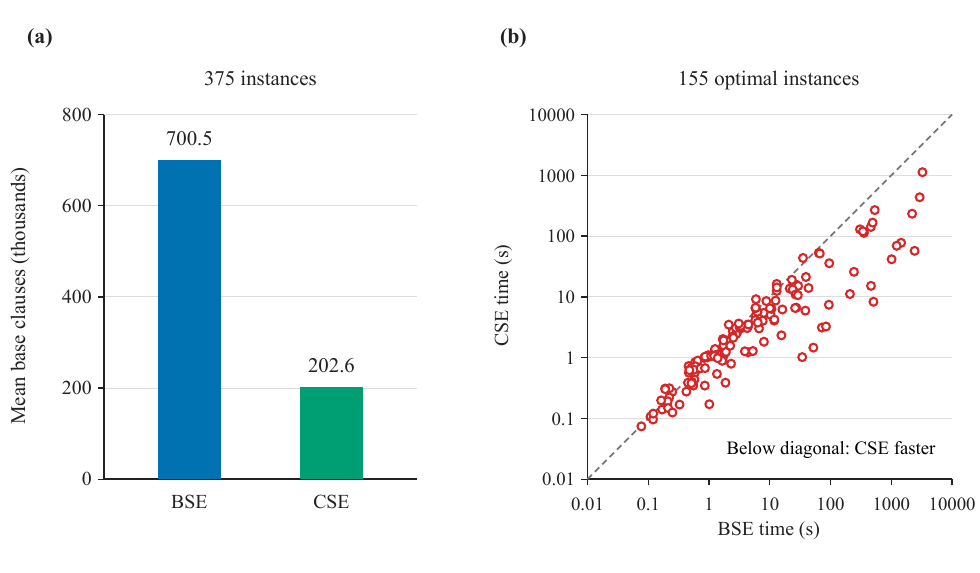}
\caption{BSE--CSE comparison.}
\label{fig:representation-ablation}
\end{figure}

Both SAT variants use the same incremental optimizer and experimental controls specified in Subsection~\ref{sec:experimental-setup}. CSE achieves a lower peak than BSE on 43 instances, matches BSE on 321, and yields a higher peak on 11. On each of the 11 instances where BSE outperforms CSE, the peak differs by one unit. Similarly, \MSLS{} achieves a peak one unit lower than CSE on four instances.
Figure~\ref{fig:representation-ablation}(a) shows that CSE reduces the average number of base-CNF clauses by 71.1\% (202,627 versus 700,453), while using more variables (4,333 versus 2,972) (Table~\ref{tab:encoding-size-summary}). This substantial reduction is achieved by introducing two additional cumulative variables, which enable CSE to replace the pairwise constraints in BSE with a more compact encoding.

CSE certifies all 155 instances certified by BSE and adds four proofs. 
Figure~\ref{fig:representation-ablation}(b) plots those 155 paired times on equal logarithmic scales. Each point represents one instance, and points below the equal-time diagonal favor CSE. The 81.4\% figure is a reduction of summed times on this subset, rather than a mean of per-run speedups or a full-benchmark reduction.
Of the full-benchmark charged-time reduction, 61.0\% comes from these 155 instances and 39.0\% from the four additional proofs; the other 216 contribute zero because both variants are charged 3600~s (Table~\ref{tab:charged-time-summary}). CSE has lower charged time than BSE on 122 instances, higher time on 36, and ties on 217; its base-instance mean is lower for 14 of 17 base instances. Counts use recorded values before rounding and include ties when neither method proves optimality.

\FloatBarrier
\begingroup
\clubpenalty=10000
Overall, CSE can support production-line design decisions under prescribed cycle times and resource limits by providing feasible schedules with lower peak power. An optimality certificate confirms that no better peak exists within the model, whereas an uncertified solution remains feasible but may admit further improvement. These implications are based on the studied benchmark and assume fixed task durations and power demands and the specified line architecture.
\par\endgroup
\FloatBarrier

\section{Conclusion}\label{sec:conclusion}

This paper proposed SAT-based optimization with incremental SAT solving for PSCALB3PM, an integrated problem combining assembly-line balancing, cyclic scheduling, resource configuration, and peak-power minimization. The proposed encoding extends our previous SALB3PM encoding by incorporating constraints to model workstations with multiple resources, cyclic scheduling, and resource limits.
We develop two SAT representations, the Basic SAT Encoding (BSE) and the Compact SAT Encoding (CSE), and integrate them with an incremental optimization strategy that progressively tightens the peak-power bound while retaining the solver's learned clauses.

Experiments on 375 benchmark instances with up to three parallel resources per workstation demonstrate the effectiveness of the proposed SAT approach. Both SAT encodings find feasible solutions for all instances and certify 155 and 159 optima, respectively, compared with 346 feasible solutions and 126 optimality certificates obtained by the reference ILP approach. Within 60 seconds, BSE and CSE certify 132 and 144 optima, respectively, compared with 66 for ILP. CSE also reduces the mean proof-oriented charged time by 3.2\% relative to BSE while establishing four additional optimality certificates. Compared with the reference matheuristic, the SAT approaches achieve lower peak power on substantially more instances, although the matheuristic provides better solutions on a small number of cases. Overall, these results show that SAT is a competitive exact approach to PSCALB3PM, providing broader feasibility coverage and stronger optimality certification than the previously proposed ILP and matheuristic methods on the tested benchmarks.

The results establish SAT as a strong baseline for future research on PSCALB3PM and motivate further investigation of improved encodings, optimization strategies, and hybrid SAT--matheuristic approaches. Future work should also evaluate larger and more diverse benchmark sets, assess the robustness of stochastic methods across repeated runs, and extend the model to settings with variable power profiles or heterogeneous resources.

\section*{Generative AI statement}
OpenAI ChatGPT and OpenAI Codex, using the GPT-6 Astra model, were used to assist with language revisions and provide some support during implementation. The authors reviewed and verified all AI-assisted material and retain full responsibility for the content, accuracy, and integrity of the work, as well as for the experimental implementation, results, and their verification.

\section*{Author contributions}
\textbf{Phong Nguyen Chi:} Conceptualization, Methodology, Formal analysis, Investigation (data collection and experimental evaluation), Software (algorithm implementation), Data curation, Visualization, Writing -- original draft, Writing -- review \& editing.

\textbf{Khanh To Van:} Conceptualization, Investigation, Methodology, Formal analysis, Supervision, Writing -- original draft, Writing -- review \& editing.

\section*{Disclosure statement}
The authors report there are no competing interests to declare.

\section*{Funding}
The authors received no specific grant from any funding agency in the public, commercial or not-for-profit sectors for this research.

\clearpage
\appendix
\renewcommand{\thetable}{\thesection\arabic{table}}
\renewcommand{\theHtable}{appendix.\thetable}
\setcounter{table}{0}

\begin{landscape}
\section{Detailed Experimental Results}\label{app:comparison}

Table~\ref{tab:method-comparison} provides a detailed experimental comparison of the four methods, grouped by base instance and resource limit.
Each labeled base instance is identified
by its underlying name and $(n,m,c)$ parameters. All gaps use the common inherited
BK defined in Eq.~\eqref{eq:eval-bk}. Mean gaps use feasible runs, whereas feas
and opt use all instances in each group; all three are percentages.
The time column reports mean charged seconds following Eq.~\eqref{eq:eval-time};
for \MSLS{}, opt is "--" and time is 3600~s.

The Total row aggregates the $(3m+3)$ \PSCALB{} instances derived from each base instance, while the "Overall" row summarizes all 375 instances. Boldface indicates a uniquely best value among the displayed results; tied best values are not boldfaced. The notation "n/a" indicates an undefined gap.

\begingroup
\footnotesize
\setlength{\LTcapwidth}{\linewidth}
\setlength{\tabcolsep}{1.4pt}
\renewcommand{\arraystretch}{1.08}
\setlength{\aboverulesep}{1.5pt}
\setlength{\belowrulesep}{1.5pt}
\newcommand{\metricheaders}{%
\multicolumn{1}{c}{$\%BK$} & \multicolumn{1}{c}{feas} &
\multicolumn{1}{c}{opt} & \multicolumn{1}{c}{time}}
\newcommand{\resultquad}[8]{%
\multirow{4}{*}{#1} & \multirow{4}{*}{#2} & \multirow{4}{*}{#3} & \multirow{4}{*}{#4} & BSE & #5\\*
& & & & \MPSC{} & #6\\*
& & & & \MSLS{} & #7\\*
& & & & CSE & #8\\
\cmidrule(lr){1-21}}
\begin{longtable}{@{}lrrr@{\hspace{4pt}}l@{\hspace{5pt}}
S[table-format=3.3]S[table-format=3.2]S[table-format=3.2]S[table-format=4.1]
@{\hspace{7pt}}
S[table-format=3.3]S[table-format=3.2]S[table-format=3.2]S[table-format=4.1]
@{\hspace{7pt}}
S[table-format=3.3]S[table-format=3.2]S[table-format=3.2]S[table-format=4.1]
@{\hspace{7pt}}
S[table-format=3.3]S[table-format=3.2]S[table-format=3.2]S[table-format=4.1]@{}}
\caption{Results of four methods: \MPSC{}, \MSLS{}, BSE, and CSE.}
\label{tab:method-comparison}\\
\toprule
\multirow{2}{*}{\shortstack[l]{Base\\instance}} & \multirow{2}{*}{$n$} & \multirow{2}{*}{$m$} &
\multirow{2}{*}{$c$} & \multirow{2}{*}{Method} &
\multicolumn{4}{c}{$r_{\max}=1$} & \multicolumn{4}{c}{$r_{\max}=2$} &
\multicolumn{4}{c}{$r_{\max}=3$} & \multicolumn{4}{c}{Total}\\
\cmidrule(lr){6-9}\cmidrule(lr){10-13}\cmidrule(lr){14-17}\cmidrule(l){18-21}
& & & & & \metricheaders & \metricheaders & \metricheaders & \metricheaders\\
\midrule
\endfirsthead
\caption[]{Results of four methods: \MPSC{}, \MSLS{}, BSE, and CSE (continued).}\\
\toprule
\multirow{2}{*}{\shortstack[l]{Base\\instance}} & \multirow{2}{*}{$n$} & \multirow{2}{*}{$m$} &
\multirow{2}{*}{$c$} & \multirow{2}{*}{Method} &
\multicolumn{4}{c}{$r_{\max}=1$} & \multicolumn{4}{c}{$r_{\max}=2$} &
\multicolumn{4}{c}{$r_{\max}=3$} & \multicolumn{4}{c}{Total}\\
\cmidrule(lr){6-9}\cmidrule(lr){10-13}\cmidrule(lr){14-17}\cmidrule(l){18-21}
& & & & & \metricheaders & \metricheaders & \metricheaders & \metricheaders\\
\midrule
\endhead
\midrule
\multicolumn{21}{r}{\footnotesize Continued on the next landscape page}\\
\endfoot
\bottomrule
\endlastfoot
\resultquad{bowman-1}{8}{5}{20}{0.000 & 100.00 & 100.00 & 0.2 & 0.000 & 100.00 & 100.00 & 7.0 & 0.000 & 100.00 & 100.00 & 27.6 & 0.000 & 100.00 & 100.00 & 19.2}{0.000 & 100.00 & 100.00 & 0.2 & 0.000 & 100.00 & 100.00 & 113.9 & 0.000 & 100.00 & 100.00 & 677.4 & 0.000 & 100.00 & 100.00 & 451.9}{0.000 & 100.00 & \tabmissing & 3600.0 & 0.000 & 100.00 & \tabmissing & 3600.0 & 0.000 & 100.00 & \tabmissing & 3600.0 & 0.000 & 100.00 & \tabmissing & 3600.0}{0.000 & 100.00 & 100.00 & 0.3 & 0.000 & 100.00 & 100.00 & \bfseries 5.1 & 0.000 & 100.00 & 100.00 & \bfseries 13.0 & 0.000 & 100.00 & 100.00 & \bfseries 9.7}
\resultquad{buxey-1}{29}{14}{25}{0.000 & 100.00 & 100.00 & \bfseries 35.2 & 0.107 & 100.00 & 0.00 & 3600.0 & 0.266 & 100.00 & 0.00 & 3600.0 & 0.207 & 100.00 & 2.22 & \bfseries 3520.8}{0.000 & 100.00 & 100.00 & 3200.5 & 5.492 & 100.00 & 0.00 & 3600.0 & 6.593 & 89.66 & 0.00 & 3600.0 & 6.043 & 93.33 & 2.22 & 3591.1}{\tabna & 0.00 & \tabmissing & 3600.0 & 0.595 & 93.33 & \tabmissing & 3600.0 & 0.421 & 100.00 & \tabmissing & 3600.0 & 0.478 & 95.56 & \tabmissing & 3600.0}{0.000 & 100.00 & 100.00 & 43.7 & \bfseries 0.043 & 100.00 & 0.00 & 3600.0 & 0.266 & 100.00 & 0.00 & 3600.0 & \bfseries 0.186 & 100.00 & 2.22 & 3521.0}
\resultquad{buxey-2}{29}{7}{47}{0.000 & 100.00 & 100.00 & 10.3 & 0.373 & 100.00 & 0.00 & 3600.0 & 0.478 & 100.00 & 0.00 & 3600.0 & 0.423 & 100.00 & 4.17 & 3450.4}{5.978 & 100.00 & 0.00 & 3600.0 & 15.483 & 87.50 & 0.00 & 3600.0 & 16.826 & 66.67 & 0.00 & 3600.0 & 15.701 & 75.00 & 0.00 & 3600.0}{\tabna & 0.00 & \tabmissing & 3600.0 & 5.397 & 100.00 & \tabmissing & 3600.0 & 2.546 & 100.00 & \tabmissing & 3600.0 & 3.537 & 95.83 & \tabmissing & 3600.0}{0.000 & 100.00 & 100.00 & \bfseries 5.0 & \bfseries 0.000 & 100.00 & 0.00 & 3600.0 & \bfseries 0.080 & 100.00 & 0.00 & 3600.0 & \bfseries 0.050 & 100.00 & 4.17 & \bfseries 3450.2}
\resultquad{jackson-1}{11}{8}{7}{0.000 & 100.00 & 100.00 & 0.2 & 0.000 & 100.00 & 100.00 & \bfseries 2.9 & 0.000 & 100.00 & 100.00 & 6.8 & 0.000 & 100.00 & 100.00 & 5.3}{0.000 & 100.00 & 100.00 & \bfseries 0.1 & 0.140 & 100.00 & 88.89 & 556.3 & 0.075 & 100.00 & 82.35 & 1721.5 & 0.094 & 100.00 & 85.19 & 1269.4}{0.000 & 100.00 & \tabmissing & 3600.0 & 0.000 & 100.00 & \tabmissing & 3600.0 & 0.000 & 100.00 & \tabmissing & 3600.0 & 0.000 & 100.00 & \tabmissing & 3600.0}{0.000 & 100.00 & 100.00 & 0.2 & 0.000 & 100.00 & 100.00 & 3.1 & 0.000 & 100.00 & 100.00 & \bfseries 6.4 & 0.000 & 100.00 & 100.00 & \bfseries 5.1}
\resultquad{jackson-2}{11}{3}{21}{0.000 & 100.00 & 100.00 & 0.3 & 0.000 & 100.00 & 100.00 & 41.0 & 0.000 & 100.00 & 100.00 & 295.3 & 0.000 & 100.00 & 100.00 & 185.9}{0.000 & 100.00 & 100.00 & 5.3 & 0.000 & 100.00 & 25.00 & 2802.2 & 0.000 & 100.00 & 14.29 & 3465.0 & 0.000 & 100.00 & 25.00 & 2955.8}{0.000 & 100.00 & \tabmissing & 3600.0 & 0.000 & 100.00 & \tabmissing & 3600.0 & 0.000 & 100.00 & \tabmissing & 3600.0 & 0.000 & 100.00 & \tabmissing & 3600.0}{0.000 & 100.00 & 100.00 & \bfseries 0.2 & 0.000 & 100.00 & 100.00 & \bfseries 29.1 & 0.000 & 100.00 & 100.00 & \bfseries 101.2 & 0.000 & 100.00 & 100.00 & \bfseries 68.8}
\resultquad{jaeschke-1}{9}{8}{6}{0.000 & 100.00 & 100.00 & 0.1 & 0.000 & 100.00 & 100.00 & \bfseries 0.9 & 0.000 & 100.00 & 100.00 & 1.7 & 0.000 & 100.00 & 100.00 & 1.3}{0.000 & 100.00 & 100.00 & 0.1 & 0.000 & 100.00 & 100.00 & 4.4 & 0.000 & 100.00 & 100.00 & 17.3 & 0.000 & 100.00 & 100.00 & 12.3}{0.000 & 100.00 & \tabmissing & 3600.0 & 0.000 & 100.00 & \tabmissing & 3600.0 & 0.000 & 100.00 & \tabmissing & 3600.0 & 0.000 & 100.00 & \tabmissing & 3600.0}{0.000 & 100.00 & 100.00 & 0.1 & 0.000 & 100.00 & 100.00 & 1.0 & 0.000 & 100.00 & 100.00 & \bfseries 1.5 & 0.000 & 100.00 & 100.00 & 1.3}
\resultquad{jaeschke-2}{9}{3}{18}{0.000 & 100.00 & 100.00 & 0.1 & 0.000 & 100.00 & 100.00 & 1.5 & 0.000 & 100.00 & 100.00 & 7.1 & 0.000 & 100.00 & 100.00 & 4.7}{0.000 & 100.00 & 100.00 & 0.6 & 0.000 & 100.00 & 100.00 & 74.2 & 0.000 & 100.00 & 100.00 & 286.1 & 0.000 & 100.00 & 100.00 & 191.7}{0.000 & 100.00 & \tabmissing & 3600.0 & 0.000 & 100.00 & \tabmissing & 3600.0 & 0.000 & 100.00 & \tabmissing & 3600.0 & 0.000 & 100.00 & \tabmissing & 3600.0}{0.000 & 100.00 & 100.00 & 0.1 & 0.000 & 100.00 & 100.00 & \bfseries 1.0 & 0.000 & 100.00 & 100.00 & \bfseries 3.3 & 0.000 & 100.00 & 100.00 & \bfseries 2.3}
\resultquad{mansoor-1}{11}{4}{48}{0.000 & 100.00 & 100.00 & 0.6 & 0.000 & 100.00 & 100.00 & 1548.3 & 0.000 & 100.00 & 22.22 & 2920.4 & 0.000 & 100.00 & 53.33 & 2268.4}{0.000 & 100.00 & 100.00 & 1.1 & 0.000 & 100.00 & 20.00 & 3171.1 & 0.000 & 100.00 & 0.00 & 3600.0 & 0.000 & 100.00 & 13.33 & 3217.1}{0.000 & 100.00 & \tabmissing & 3600.0 & 0.000 & 100.00 & \tabmissing & 3600.0 & 0.000 & 100.00 & \tabmissing & 3600.0 & 0.000 & 100.00 & \tabmissing & 3600.0}{0.000 & 100.00 & 100.00 & 0.6 & 0.000 & 100.00 & 100.00 & \bfseries 330.1 & 0.000 & 100.00 & \bfseries 33.33 & \bfseries 2631.3 & 0.000 & 100.00 & \bfseries 60.00 & \bfseries 1688.9}
\resultquad{mansoor-2}{11}{2}{94}{0.000 & 100.00 & 100.00 & 1.0 & 0.000 & 100.00 & 100.00 & 193.9 & 0.000 & 100.00 & 60.00 & 2036.4 & 0.000 & 100.00 & 77.78 & 1196.1}{0.000 & 100.00 & 100.00 & 2.7 & 0.000 & 100.00 & 100.00 & 1632.4 & 0.000 & 100.00 & 20.00 & 3153.9 & 0.000 & 100.00 & 55.56 & 2296.6}{0.000 & 100.00 & \tabmissing & 3600.0 & 0.000 & 100.00 & \tabmissing & 3600.0 & 0.000 & 100.00 & \tabmissing & 3600.0 & 0.000 & 100.00 & \tabmissing & 3600.0}{0.000 & 100.00 & 100.00 & \bfseries 0.2 & 0.000 & 100.00 & 100.00 & \bfseries 6.5 & 0.000 & 100.00 & \bfseries 100.00 & \bfseries 318.6 & 0.000 & 100.00 & \bfseries 100.00 & \bfseries 179.2}
\resultquad{mertens-1}{7}{6}{6}{0.000 & 100.00 & 100.00 & 0.1 & 0.000 & 100.00 & 100.00 & 0.2 & 0.000 & 100.00 & 100.00 & \bfseries 0.5 & 0.000 & 100.00 & 100.00 & 0.4}{0.000 & 100.00 & 100.00 & 0.1 & 0.000 & 100.00 & 100.00 & 0.7 & 0.000 & 100.00 & 100.00 & 1.2 & 0.000 & 100.00 & 100.00 & 1.0}{0.000 & 100.00 & \tabmissing & 3600.0 & 0.000 & 100.00 & \tabmissing & 3600.0 & 0.000 & 100.00 & \tabmissing & 3600.0 & 0.000 & 100.00 & \tabmissing & 3600.0}{0.000 & 100.00 & 100.00 & 0.1 & 0.000 & 100.00 & 100.00 & 0.2 & 0.000 & 100.00 & 100.00 & 0.6 & 0.000 & 100.00 & 100.00 & 0.4}
\resultquad{mertens-2}{7}{2}{18}{0.000 & 100.00 & 100.00 & 0.1 & 0.000 & 100.00 & 100.00 & 0.4 & 0.000 & 100.00 & 100.00 & 1.3 & 0.000 & 100.00 & 100.00 & 0.9}{0.000 & 100.00 & 100.00 & 0.2 & 0.000 & 100.00 & 100.00 & 2.7 & 0.000 & 100.00 & 100.00 & 4.8 & 0.000 & 100.00 & 100.00 & 3.6}{0.000 & 100.00 & \tabmissing & 3600.0 & 0.000 & 100.00 & \tabmissing & 3600.0 & 0.000 & 100.00 & \tabmissing & 3600.0 & 0.000 & 100.00 & \tabmissing & 3600.0}{0.000 & 100.00 & 100.00 & 0.1 & 0.000 & 100.00 & 100.00 & \bfseries 0.3 & 0.000 & 100.00 & 100.00 & \bfseries 1.0 & 0.000 & 100.00 & 100.00 & \bfseries 0.6}
\resultquad{mitchell-1}{21}{8}{14}{0.000 & 100.00 & 100.00 & 0.9 & 0.000 & 100.00 & 11.11 & 3259.4 & 0.110 & 100.00 & 0.00 & 3600.0 & 0.069 & 100.00 & 7.41 & 3353.2}{0.000 & 100.00 & 100.00 & 3.7 & 0.674 & 100.00 & 0.00 & 3600.0 & 0.851 & 100.00 & 0.00 & 3600.0 & 0.761 & 100.00 & 3.70 & 3466.8}{\tabna & 0.00 & \tabmissing & 3600.0 & 0.569 & 100.00 & \tabmissing & 3600.0 & 0.247 & 100.00 & \tabmissing & 3600.0 & 0.359 & 96.30 & \tabmissing & 3600.0}{0.000 & 100.00 & 100.00 & \bfseries 0.7 & 0.000 & 100.00 & 11.11 & \bfseries 3229.8 & \bfseries 0.082 & 100.00 & 0.00 & 3600.0 & \bfseries 0.052 & 100.00 & 7.41 & \bfseries 3343.3}
\resultquad{mitchell-2}{21}{3}{39}{0.000 & 100.00 & 100.00 & 2.3 & 0.000 & 100.00 & 50.00 & 2359.7 & 0.362 & 100.00 & 14.29 & 3115.5 & 0.211 & 100.00 & 33.33 & 2604.1}{0.000 & 100.00 & 100.00 & 562.6 & 1.258 & 100.00 & 0.00 & 3600.0 & 1.625 & 100.00 & 0.00 & 3600.0 & 1.367 & 100.00 & 8.33 & 3346.9}{0.000 & 100.00 & \tabmissing & 3600.0 & 2.203 & 100.00 & \tabmissing & 3600.0 & 1.987 & 100.00 & \tabmissing & 3600.0 & 1.893 & 100.00 & \tabmissing & 3600.0}{0.000 & 100.00 & 100.00 & \bfseries 0.8 & 0.000 & 100.00 & 50.00 & \bfseries 1859.9 & \bfseries 0.181 & 100.00 & \bfseries 28.57 & \bfseries 2655.2 & \bfseries 0.105 & 100.00 & \bfseries 41.67 & \bfseries 2168.9}
\resultquad{roszieg-1}{25}{10}{14}{0.000 & 100.00 & 100.00 & 2.2 & 0.000 & 100.00 & 0.00 & 3600.0 & 0.000 & 100.00 & 0.00 & 3600.0 & 0.000 & 100.00 & 3.03 & 3491.0}{0.000 & 100.00 & 100.00 & 124.4 & 0.767 & 100.00 & 0.00 & 3600.0 & 0.623 & 100.00 & 0.00 & 3600.0 & 0.652 & 100.00 & 3.03 & 3494.7}{\tabna & 0.00 & \tabmissing & 3600.0 & 0.345 & 100.00 & \tabmissing & 3600.0 & 0.120 & 100.00 & \tabmissing & 3600.0 & 0.197 & 96.97 & \tabmissing & 3600.0}{0.000 & 100.00 & 100.00 & \bfseries 1.6 & 0.000 & 100.00 & 0.00 & 3600.0 & 0.000 & 100.00 & 0.00 & 3600.0 & 0.000 & 100.00 & 3.03 & 3491.0}
\resultquad{roszieg-2}{25}{4}{32}{0.000 & 100.00 & 100.00 & 3.9 & 0.192 & 100.00 & 20.00 & 2928.6 & 0.427 & 100.00 & 11.11 & 3336.3 & 0.321 & 100.00 & 20.00 & 2978.2}{0.000 & 100.00 & 100.00 & 496.9 & 3.706 & 100.00 & 0.00 & 3600.0 & 3.664 & 100.00 & 0.00 & 3600.0 & 3.434 & 100.00 & 6.67 & 3393.1}{\tabna & 0.00 & \tabmissing & 3600.0 & 3.555 & 100.00 & \tabmissing & 3600.0 & 2.745 & 100.00 & \tabmissing & 3600.0 & 3.034 & 93.33 & \tabmissing & 3600.0}{0.000 & 100.00 & 100.00 & \bfseries 1.3 & \bfseries 0.000 & 100.00 & 20.00 & \bfseries 2885.2 & \bfseries 0.321 & 100.00 & 11.11 & \bfseries 3207.7 & \bfseries 0.192 & 100.00 & 20.00 & \bfseries 2886.4}
\resultquad{sawyer-1}{30}{14}{25}{0.000 & 100.00 & 100.00 & \bfseries 5.8 & \bfseries 0.049 & 100.00 & 0.00 & 3600.0 & 0.304 & 100.00 & 0.00 & 3600.0 & 0.212 & 100.00 & 2.22 & 3520.1}{0.000 & 100.00 & 100.00 & 173.4 & 1.471 & 6.67 & 0.00 & 3600.0 & 10.445 & 100.00 & 0.00 & 3600.0 & 9.819 & 68.89 & 2.22 & 3523.9}{\tabna & 0.00 & \tabmissing & 3600.0 & 1.173 & 100.00 & \tabmissing & 3600.0 & 0.393 & 100.00 & \tabmissing & 3600.0 & 0.659 & 97.78 & \tabmissing & 3600.0}{0.000 & 100.00 & 100.00 & 6.6 & 0.098 & 100.00 & 0.00 & 3600.0 & \bfseries 0.266 & 100.00 & 0.00 & 3600.0 & \bfseries 0.204 & 100.00 & 2.22 & 3520.1}
\resultquad{sawyer-2}{30}{7}{47}{0.000 & 100.00 & 100.00 & 11.9 & 0.429 & 100.00 & 0.00 & 3600.0 & 0.825 & 100.00 & 0.00 & 3600.0 & 0.659 & 100.00 & 4.17 & 3450.5}{0.000 & 100.00 & 0.00 & 3600.0 & 18.408 & 50.00 & 0.00 & 3600.0 & 11.222 & 86.67 & 0.00 & 3600.0 & 12.195 & 75.00 & 0.00 & 3600.0}{\tabna & 0.00 & \tabmissing & 3600.0 & 4.523 & 100.00 & \tabmissing & 3600.0 & 2.699 & 100.00 & \tabmissing & 3600.0 & 3.333 & 95.83 & \tabmissing & 3600.0}{0.000 & 100.00 & 100.00 & \bfseries 4.2 & \bfseries 0.086 & 100.00 & 0.00 & 3600.0 & \bfseries 0.551 & 100.00 & 0.00 & 3600.0 & \bfseries 0.373 & 100.00 & 4.17 & \bfseries 3450.2}
\resultquad{\textbf{Overall (375)}}{}{}{}{0.000 & 100.00 & 100.00 & 4.4 & 0.078 & 100.00 & 43.20 & 2137.5 & 0.190 & 100.00 & 36.05 & 2336.8 & 0.144 & 100.00 & 41.33 & 2164.6}{0.352 & 100.00 & 88.24 & 692.5 & 2.879 & 84.80 & 33.60 & 2464.0 & 3.864 & 95.71 & 29.61 & 2672.4 & 3.390 & 92.27 & 33.60 & 2513.2}{0.000 & 58.82 & \tabmissing & 3600.0 & 1.135 & 99.20 & \tabmissing & 3600.0 & 0.633 & 100.00 & \tabmissing & 3600.0 & 0.786 & 97.87 & \tabmissing & 3600.0}{0.000 & 100.00 & 100.00 & \bfseries 3.9 & \bfseries 0.022 & 100.00 & 43.20 & \bfseries 2064.0 & \bfseries 0.131 & 100.00 & \bfseries 37.77 & \bfseries 2263.3 & \bfseries 0.089 & 100.00 & \bfseries 42.40 & \bfseries 2094.4}

\end{longtable}
\endgroup
\end{landscape}


Table~\ref{tab:proof-time} reports cumulative counts of instances proved optimal,
out of 375, at selected time thresholds from Figure~\ref{fig:proof-profile}.
Counts use recorded proof time and the
OPTIMAL status of each run.

\begin{table}[H]
\centering
\let\proofcaptionfont\tablecaptionfont
\renewcommand{\tablecaptionfont}{\proofcaptionfont\centering}
\caption{Optimality Results at Selected Time Limits.}
\label{tab:proof-time}
\begin{tabular}{@{}lrrrrrr@{}}
\toprule
Method & 1 s & 10 s & 60 s & 600 s & 1800 s & 3600 s\\
\midrule
BSE & 41 & 106 & 132 & 148 & 151 & 155\\
CSE & \textbf{45} & \textbf{123} & \textbf{144} & \textbf{156} & \textbf{158} & \textbf{159}\\
\MPSC{} & 16 & 49 & 66 & 105 & 119 & 126\\
\bottomrule

\end{tabular}
\end{table}

\clearpage
\section{Encoding Size and Charged-Time Reduction}
\label{app:encoding-summary}
\setcounter{table}{0}

Table~\ref{tab:encoding-size-summary} reports the variables and clauses in the base formula
in conjunctive normal form (CNF), before the incremental peak-indicator constraints.
Every instance has equal weight within its resource-limit group. Clause reduction
is $100(1-\overline{C}_{\mathrm{CSE}}/\overline{C}_{\mathrm{BSE}})$,
where $\overline{C}$ is the mean clause count in the group. CSE uses more base variables
but 71.1\% fewer base clauses on average; the clause reduction rises
from 42.8\% to 73.4\% across the three $r_{\max}$ groups. These counts
exclude the subsequently added peak-bound encodings and do not identify
which component causes the observed time difference.

\begin{table}[htbp]
\centering
\caption{Mean base-CNF size.}
\label{tab:encoding-size-summary}
\small
\setlength{\tabcolsep}{5pt}
\begin{tabular}{@{}lrrrrrr@{}}
\toprule
& & \multicolumn{2}{c}{BSE} & \multicolumn{2}{c}{CSE} & \\
\cmidrule(lr){3-4}\cmidrule(lr){5-6}
$r_{\max}$ & Instances & variables & clauses & variables & clauses &
reduction (\%)\\
\midrule
1 & 17 & 1,061 & 94,406 & 1,525 & 54,014 & 42.8 \\
2 & 125 & 2,291 & 409,894 & 3,336 & 154,531 & 62.3 \\
3 & 233 & 3,477 & 900,550 & 5,073 & 239,273 & 73.4 \\
Overall & 375 & 2,972 & 700,453 & 4,333 & 202,627 & 71.1 \\
\bottomrule
\end{tabular}
\end{table}

For any group $\mathcal G$, its contribution to the total charged-time
reduction is
\[
D_{\mathcal G}=\sum_{i\in\mathcal G}
\bigl(\widetilde t_{\mathrm{BSE},i}
      -\widetilde t_{\mathrm{CSE},i}\bigr).
\]
Table~\ref{tab:charged-time-summary} partitions all 375 instances
according to certification status. The 155 jointly certified cases
contribute 16069.095~s, or 61.0\% of the total reduction. The four
additional proofs contribute 10261.663~s, or 39.0\%. On the remaining
216 instances, both methods are charged 3600~s, so their
contribution is zero irrespective of when incumbents were found.
Dividing the total reduction of 26330.758~s by 375 gives a
70.215~s reduction in mean charged time, underlying the 70.2~s
difference between the rounded means in Section~\ref{sec:proof-time}.
This decomposition describes the chosen metric; it is not a statistical
test or a causal attribution to individual encoding components.
The table reports time sums in seconds, not per-instance means.
Share is $100D_{\mathcal G}/D_I$, where $I$ is the full set of 375 instances.

\begin{table}[htbp]
\centering
\caption{Charged-time reduction.}
\label{tab:charged-time-summary}
\small
\setlength{\tabcolsep}{3.5pt}
\begin{tabular}{@{}lrrrrr@{}}
\toprule
Certification group & Instances & BSE total & CSE total & Reduction & Share (\%)\\
\midrule
Both certified & 155 & 19,741.842 & 3,672.747 & 16,069.095 & 61.0 \\
CSE only certified & 4 & 14,400.000 & 4,138.337 & 10,261.663 & 39.0 \\
Neither certified & 216 & 777,600.000 & 777,600.000 & 0.000 & 0.0 \\
\midrule
Overall & 375 & 811,741.842 & 785,411.084 & 26,330.758 & 100.0 \\
\bottomrule
\end{tabular}
\end{table}

\end{document}